\documentclass[superscriptaddress,nofootinbib,twocolumn,pre]{revtex4-1}

\usepackage{amsmath,parskip,graphicx,amssymb,acro,array}
\usepackage{siunitx}  
\usepackage[titletoc]{appendix}
\usepackage[normalem]{ulem}
\usepackage{color}

\newcommand{\correction}[1]{#1}
\newcommand{\cutout}[1]{}

\begin{document}

\begin{abstract}
We present a theoretical framework based on an extension of dynamical density functional theory (DDFT) for describing the structure and dynamics of cells in living tissues and tumours. DDFT is a microscopic statistical mechanical theory for the time evolution of the density distribution of interacting many-particle systems. The theory accounts for cell pair-interactions, different cell types, phenotypes and cell birth and death processes (including cell division), in order to provide a biophysically consistent description of processes bridging across the scales, including describing the tissue structure down to the level of the individual cells. Analysis of the model is presented for a single species and a two-species cases, the latter \correction{aimed at} describing competition between tumour and healthy cells. In suitable parameter regimes, model results are consistent with biological observations. Of particular note, divergent tumour growth behaviour, mirroring metastatic and benign growth characteristics, are shown to be dependent on the cell pair-interaction parameters.

\end{abstract}


\title{\correction{Dynamical density functional theory based modelling of tissue dynamics: application to tumour growth\\
\cutout{ Dynamical density functional theory based modelling of tumour growth}}}
\author{Hayder M. Al-Saedi}
\affiliation{Department of Mathematical Sciences, Loughborough University, Loughborough, LE11 3TU, UK}
\affiliation{Department of Mathematical Sciences, Baghdad University, Baghdad, Iraq}
\author{Andrew J. Archer}
\affiliation{Department of Mathematical Sciences, Loughborough University, Loughborough, LE11 3TU, UK}
\author{John Ward}
\affiliation{Department of Mathematical Sciences, Loughborough University, Loughborough, LE11 3TU, UK}
\date{\today}

\maketitle



\section{Introduction}

One of the characteristics of biological systems is their ability to produce and sustain spatiotemporal patterns -- i.e.\ structure formation. Cancer is a disease that may be viewed as a complex system whose dynamics and growth results from nonlinear processes coupled across a wide range of spatiotemporal scales. Cancer is recognised as one of the major causes of premature death, soon to overtake heart disease as the leading cause in the developed nations \cite{byrne2000using}. At current rates, in the USA a third of women and half of men will develop a cancer at some point in their life \cite{siegel2012cancer}. Though significant progress has been made in cancer treatment in recent decades, much research is still required in order to control all forms of the disease.

The human body is made up of order $10^{13}$ cells. Genetic mutations are frequent, but most affected cells die by apoptosis and are removed by the immune system. However, a few may escape the regulatory process to produce an abnormally growing colony that in time recruits its own vascular system (via angiogenesis) and form a cancer. Tumour growth varies and solid tumours can be classified as either benign or malignant \cite{weinberg2013biology}. The former are localised, but their continued growth can cause damage to neighbouring healthy tissues from the mechanical forces applied. Whilst most tumours are initially benign, malignancy can develop, whereby individual cells are able to escape the main tumour mass (metastasis) and colonise elsewhere in the body; it is these cells that give rise to the greatest clinical concern. 

Much work has gone into developing mathematical models of cancers. Of particular interest here is the spatiotemporal dynamics, which can be described e.g.\ using continuum, discrete and hybrid models. Continuum approaches usually result in a system of coupled partial differential equations and have been used to describe avascular growth \cite{burton1966rate,greenspan1972models,sutherland1971growth,glass1973instability,bullough1965mitotic,ward1997mathematical,lowengrub2009nonlinear}, vascular growth \cite{adam1986simplified,mcelwain1979model,burton1966rate,breward2003multiphase,orme1996mathematical}, angiogenesis \cite{byrne1995growth,kerbel2000tumor,viallard2017tumor} and treatment \cite{kim2008pde,miller2016cancer,ward2003mathematical}. Most of these consider the overall growth as being dependent on nutrient(s) that diffuses in from the outside, whilst more sophisticated extensions of these models treat the tumour as a poro-viscous  \cite{ruoslahti1996cancer,liotta1974quantitative,liotta1976significance} or poro-elastic \cite{byrne1995growth,adam1986simplified,adam1987mathematical,chen2014mathematical} structure. In such models the cell-cell interactions enter via coefficients in the mass conservation terms and (usually) linear constitutive relations describing the macroscale material properties of the tissue, rather than via any genuine microscale description of the interaction between cells. Of course, the advantage of such models is that they are amenable to analytical techniques and relatively small-scale computation. However, the microscopic cell-cell interactions play a crucial role in the development and function of multicellular organisms \cite{pierres2000cell}, so it is desirable to incorporate cell-cell interaction effects in the modelling. These interactions determine the structural integrity of tissue and allow cells to communicate with each other in response to changes in their micro-environment, which is essential for the survival of the cells and the host. Such communication includes that from physical contact and chemical signals, transported directly through gap junctions between cells or by passive diffusion. Some of these aspects can differ between healthy and cancer cells, so modelling these differences can be important.

Greater detail of the cell-cell interactions are routinely incorporated in discrete models for tumour growth, such as cellular-automata \cite{piotrowska2009quantitative,anderson2005hybrid,engelberg2008essential}, agent-based models \cite{drasdo2005single,galle2009single,jeon2010off} and Potts models  \cite{turner2004tamoxifen,maree2007cellular,shirinifard20093d}. In these, cells are described at a microscopic level as entities that move and respond to neighbours via a set of biologically motivated rules. Simulating the action of a group of many of these cells then gives the evolution of a tumour on the macroscale. Cellular automata models consists of a regular grid of cells, each in one of a finite number of states, such as `on' or `off'. In agent based models their actions typically follow discrete event cues or a sequential schedule of interactions, rather than simultaneously performing actions at constant time-steps, as in cellular automata models. Potts type models are able to incorporate how internal elements of the cells respond to one another based on certain characteristics that each \correction{posess} \cutout{ has} \cite{anderson2005hybrid,jeon2010off,maree2007cellular}. Though discrete models are good for incorporating the biology and physics of cell-cell interactions, they are designed for computation and are generally difficult to study analytically. 

A continuum theory that also incorporates the cell-cell interactions at a microscopic level was proposed (but not analysed) in Ref.~\cite{CHKLC2012}. The central idea is to base the model on dynamical density functional theory (DDFT) \cite{marconi1999dynamic,ArEv04,archer2004}, which is a theory for the dynamics of interacting Brownian (colloidal) particles, able to describe the time evolution of variations of the density distribution of the particles over length scales comparable with the size of the individual particles. This is the approach we extend and implement here. DDFT provides a systematic means of obtaining a continuum description of the density distribution of the cells that also incorporates a description of the microscale interactions between cells. One can solve the DDFT numerically for large enough systems to enable a macroscopic description at the population level, but perhaps more importantly is amenable to mathematical analysis (e.g.\ determination of linear stability thresholds) that gives good insight to the population collective behavior. DDFT is itself based on equilibrium density functional theory (DFT), an approach that has long been used to describe the structure of matter, be it (crystalline) solid, liquid or gas \cite{evans1979nature,Evans92,hansen2013theory}. We analyse in detail a version of the DDFT proposed in \cite{CHKLC2012} (here we specify a particular model for the interaction potential between cells) and also extend the model to describe the dynamics of systems \correction{representing} \cutout{ containing} multiple cell types, incorporating the various different pair interactions between pairs of healthy cells, between pairs of cancer cells and the cancer-healthy pair interaction. The DDFT we use is based on a DFT able to describe both the fluid and (crystalline) solid phases of soft particles. In the latter, the density distribution corresponds to a regular array of peaks, defining where the particles are located. It is in this regime, where the peaks represent the loci of cell centres, that the theory is relevant to describing the microscopic density distribution of both cancer and healthy cells, which are treated as soft particles.

This paper is laid out as follows: In section \ref{sec1} we present the DDFT for a single species of cells, perform a linear stability analysis and present some typical simulation results. In section \ref{sec61} we extend this model to describe the competition between cancer and healthy cells, and again elucidate the behaviour of the model using a linear stability analysis and simulations. Finally, in section \ref{con}, we present our conclusions.


\section{Model for a single species of cells}\label{sec1}

\subsection{Dynamical Density Functional Theory}
\label{sec:Method}

DDFT \cite{marconi1999dynamic,ArEv04,archer2004} is a theory for the spatiotemporal evolution of the ensemble average number density distribution $\rho(\textbf{r},t)$ of a system of interacting Brownian particles, where $t$ is the time and $\textbf{r}$ is the position in space. The theory shows that the dynamics is given by 
\begin {equation}\label{eq:66}
\frac{\partial \rho (\textbf{r},t)}{\partial t}=\Gamma\nabla\cdot\left[ \rho(\textbf{r},t)\nabla\left(\frac{\delta {\cal F}[\rho(\textbf{r},t]}{\delta\rho(\textbf{r},t)}\right)\right],
\end{equation}
where $\Gamma$ is a mobility coefficient and
 \begin {eqnarray} \nonumber \label{eq:8}
{\cal F}[\rho({\mathbf{r}})]&=&k_B T \int d\textbf{r} \rho({\mathbf{r}})( \ln [\Lambda^d \rho({\mathbf{r}})]-1)+{\cal F}_{ex}[\rho({\mathbf{r}})]\\
&& +\int d \textbf{r} V_{ext}(\textbf{r})\rho(\textbf{r})
\end{eqnarray}
is the Helmholtz free energy functional from equilibrium DFT \cite{hansen2013theory,evans1979nature,Evans92}. The first term in \eqref{eq:8} is the ideal gas contribution to the free energy, $d$ is the dimensionality of space, $k_B$ is Boltzmann's constant, $T$ is the temperature, $\Lambda$ is the thermal de Broglie wavelength, $V_{ext}(\textbf{r})$ is the external potential and ${\cal F}_{ex}[\rho({\mathbf{r}})]$ is the excess contribution due to the interactions between particles. In general, ${\cal F}_{ex}[\rho({\mathbf{r}})]$ is not known exactly. However, there are many different approximations which may be used \cite{Evans92,hansen2013theory}, with some being more appropriate than others, depending on the nature of the interactions between the fluid particles.

The equilibrium properties of the system are obtained by minimising the grand potential functional
\begin{equation}\label{16}
\Omega[\rho({\mathbf{r}})]={\cal F}[\rho({\mathbf{r}})]-\mu \int d{\mathbf{r}} \rho({\mathbf{r}}),
\end{equation}
where $\mu$ is the chemical potential, which is effectively the Lagrange multiplier that enforces the constraint that the average number of particles in the system is $N=\int d{\mathbf{r}} \rho({\mathbf{r}})$. Note that Eq.\ \eqref{eq:66} also enforces this constraint due to having the form of a continuity equation.

The equation of motion for each of the $N$ interacting particles (cells) that is assumed in deriving Eq.\ \eqref{eq:66} is the following over-damped Langevin equation
\begin {equation}\label{eq:1}
\frac{d\textbf{r}_i}{dt}=\Gamma \left(\textbf{F}_i^{ext}+\sum_{j=1}^{N}\textbf{F}_{ij}^{int} \right)+\sqrt{2D} \boldsymbol{\eta}_i(t),
\end{equation} 
where $\textbf{r}_i$ is the position of the centre of mass of the $i$-th particle and $D=\Gamma k_B T$ is the diffusion coefficient. This assumes no cell-cell friction; incorporating such friction would involve the inclusion of  an additional viscous drag force in the Langevin equation. The force $\textbf{F}_i^{ext}=-\nabla V_{ext}(\textbf{r}_i,t)$ is the force due to the external potential, e.g.\ due to any confining structures present, and the force $\textbf{F}_{ij}^{int}=-\nabla V_{int} ( \textbf{r}_i-\textbf{r}_j )$ is cell-cell interaction force between particles $i$ and $j$, that is assumed to be governed by the pair potential $V_{int}$ that depends on the distance between the two cells. The vector $\boldsymbol{\eta}_i(t)$ is a Gaussian random noise with components ${\eta}_i^{\alpha}(t)$ satisfying $\langle {\eta}_i^{\alpha}(t) \rangle=0$ and $\langle {\eta}_i^{\alpha}(t){\eta}_j^{\beta}(t^{'})\rangle =\delta_{ij} \delta_{\alpha \beta}(t-t^{'})$, where $\langle \cdot \rangle$ denotes a statistical average over different noise realisations, and $\alpha$, $\beta$ are coordinate indices. 

\subsection{Extension to describe living cells}\label{app}

As discussed in Ref.\ \cite{CHKLC2012}, if living cells (density $\rho(\mathbf{r},t)$) are treated as interacting Brownian particles, then an equation for the time evolution of the density of the form of Eq.\ \eqref{eq:66} is appropriate. However, since the cells can reproduce and die, there is an additional term $D^{(1)}_{BD}[\rho(\textbf{r})]$ added to the right hand side of Eq.\ \eqref{eq:66} to describe the non-conserved component of the dynamics due to birth and death (BD) processes.

As a simple model of BD, we assume that a single cell can undergo mitosis with a nutrient-dependent rate $a_m=a_m(n)$, where $n(\mathbf{r},t)$ is the local concentration of nutrient (e.g.\ dissolved O$_2$). We model cell death (apoptosis) as occurring  with a rate constant $\lambda_d$. This can be implemented as a Markov process and affects the number of cells in the population $N=N(t)$ \cite{CHKLC2012}. The nutrient is provided by the vascular system, diffuses through the system and is taken-up by the cells, and thus satisfies the reaction-diffusion equation
  \begin{equation}\label{vvv}
  \frac{\partial n(\textbf{r},t)}{\partial t}=D_n \nabla^{2}n(\textbf{r},t)+S_n f(\textbf{r})-\lambda_n\rho (\textbf{r},t) n(\textbf{r},t),
  \end{equation}
 where $D_n$ is the nutrient diffusion coefficient, $S_n$ represent the amplitude of the nutrient source, $f(\mathbf{r})$ is a function that defines where in space the nutrient source is located. Here, we consider both a uniform source $f(\mathbf{r})=1$ and a localised source in the form of Gaussian, namely 
 \begin {equation}\label{ge}
{f(\mathbf{r})}=e^{{-(x-L/2)}^{2}},
 \end {equation}
which corresponds to a source of nutrient along the line $x=\frac{L}{2}$ where $L$ is the domain width, e.g.\ due to a capillary being there. Here, $\lambda_n$ is a nutrient uptake rate constant. The term in Eq.\ \eqref{vvv} describing this process is assumed to be proportional to $n$. From the fact that the first moment of the BD process is the result of two mass action laws gives $D^{(1)}_{BD}[\rho]=a_m(n)\rho-\lambda_d \rho$, where $a_m(n)$ is a nutrient-dependent growth rate and $\lambda_d$ is a death rate constant. We assume that $a_m(n)=\lambda_m n$, where $\lambda_m$ is constant.

As a simple model for the cell-cell forces, we assume the cells interact via a soft, purely repulsive and radially symmetric pair potential
\begin{equation}\label{dafra}
V_{int}(r)= \varepsilon \exp[-(r/R)^{\cal N}],
\end{equation}
where $r$ is the distance between the centres of the cells and the parameters $\varepsilon$ and $R$ are the cell-cell interaction energy and cell radius, respectively, defining the strength and range of the potential. This is the so called generalized exponential model with exponent ${\cal N}$, or `GEM-${\cal N}$' potential \cite{likos2001effective}. Here, we set the exponent ${\cal N}=4$. Such soft potentials arise as the coarse-grained effective potential between soft polymeric macromolecules in solution 
\cite{likos2001effective,DaHa94,likos:prl:98,LBHM00,Dzubiella_2001,GHL04,MFKN05,Likos06,LBLM12}. \correction{In this study, the parameter $R$ typically represents the radius of a cell, so cells repulse each other when the distance between their centres are less than $2R$. Whilst this property of $V_{int}$ is necessary for biological relevance, longer range effects (for distances $\geq 2R$), such as cell-cell adhesion \cite{drasdo2005single,jeon2010off,shirinifard20093d},  can be straightforwardly built in to the interaction function \cite{evans1979nature,Evans92,hansen2013theory}. Note also that whilst adhesion is important for maintaining cohesion, the structure of condensed systems is dominated by the inter-particle repulsions \cite{hansen2013theory}. }

We consider this model because the bulk structure and phase behaviour of the GEM-${\cal N}$ systems are well understood in both two-dimensions (2D) and three-dimensions and also the following simple approximation for the excess free energy functional is fairly accurate and widely used \cite{likos2001effective,archer2001binary,ALE04,GAL06,MGKNL06,MoLi07,OvLi09b,CPPR12,ARK13,archer2014solidification},
\begin{equation}\label{eq:F_ex}
{\cal F}_{ex}[\rho({\mathbf{r}})]=\frac{1}{2}\int d\textbf{r}\int d\textbf{r}{'} \rho(\textbf{r})\rho(\textbf{r}{'})V_{int}(|\textbf{r}-\textbf{r}{'}|).
\end{equation}
Taking the functional derivative and then substituting the result into the extension of Eq.\ \eqref{eq:8} including the BD term described above, we obtain
\begin{align}\nonumber \label{g}
\frac{\partial \rho(\textbf{r},t)}{\partial t}&= \nabla \cdot \bigg [ \Gamma \rho(\textbf{r},t)\nabla [k_B T \ln (\Lambda^d \rho(\textbf{r},t)) \\
&~~~+\int d {\bf{r'}} \rho ({\bf{r'}},t)  V_{int}(|\mathbf{r-r}'|)]\bigg] \nonumber \\
&\mathrel{\phantom{=}} +[ \lambda_mn(\textbf{r},t)-\lambda_d] \rho (\textbf{r},t).
\end{align}
The coupled pair, Eqs.\ \eqref{g} and \eqref{vvv}, define our model for a single type of cells coupled to a source of nutrients. The parameters and their estimated values are listed in Table \ref{table:nonlin2}. See also the Appendix, where we justify the particular values we use here. For simplicity, we henceforth assume the system is 2D within a square domain of area $L^2$ with periodic boundary conditions. Thus, $\textbf{r}=(x,y)$. Two key quantities for understanding the behaviour of the system are the average cell and nutrient densities in the domain defined as
\begin {equation}\label{acd}
 \bar \rho(t)=\frac{1}{L^2}\int \int \rho(x,y,t) dx dy,
\end{equation}
\begin{equation}\label{annd}
 \bar n(t)=\frac{1}{L^2}\int \int n(x,y,t) dx dy,
 \end {equation}
respectively.

\subsection{Nondimensionalization}\label{nond5}

We now nondimensionlise the model, before performing a linear stability analysis and then presenting some typical numerical results. Writing
\begin{eqnarray}\label {l}\nonumber
 t&=& \frac{R^{2} t^{*}}{D_c},~
x=\frac{x^{*}}{R},~
y=\frac{y^{*}}{R},~
 \rho=\frac{\rho^{*}}{R^{2}},~
 n= \frac{\lambda_d n^{*}}{\lambda_m},\\
&&V_{int}(r/R)=  \varepsilon \tilde {V}_{int}(r^*),
\end{eqnarray}
where the asterisked quantities are dimensionless variables, $D_c = \Gamma k_B T$ is the dimensional coefficient of diffusion of cells and $\tilde {V}_{int}(r)=\exp(-r^{\cal N})$ is the dimensionless pair potential. We also define the dimensionless parameters
\begin{equation}\label{lamda}
  c_1=\frac {R^2 \lambda_d}{D_c}, \tilde D =\frac {D_n}{D_c}, \    \tilde S_n=\frac{R^2 S_n \lambda_m}{\lambda_d D_c},   \  \tilde \lambda_n=\frac{\lambda_n}{{D_c}}.
   \end{equation}
With these, we obtain the following nondimensional pair of coupled equations
\begin{eqnarray}\label{first}\nonumber
\frac{\partial \rho(\textbf{r},t) }{\partial t }&=&\nabla^2 \rho(\textbf{r},t)\\
&&+\nabla \cdot \bigg ( \rho (\textbf{r},t) \nabla \int d \textbf{r}{'}\rho(\textbf{r}{'},t)  \beta \varepsilon \tilde V_{int}(|\mathbf{r-r}'|)\bigg ) \nonumber \\
&& +c_1 \left[ n(\textbf{r} ,t)-1 \right] \rho (\textbf{r},t),
\end{eqnarray}
\begin{equation}\label{second}
  \frac{\partial n(\textbf{r},t)}{\partial t }=\tilde D \nabla^{2}n(\textbf{r},t)+\tilde S_n f(\textbf{r})- \tilde \lambda_n \rho (\textbf{r},t)  n(\textbf{r},t),
  \end{equation}
where we have dropped the asterisks for clarity. Note that $\beta=1/k_B T$ so that the dimensionless quantity $\beta\varepsilon$ in the integral term is the dimensionless pair interaction energy.

Our estimated values for the various dimensionless parameters in the model are listed in Table \ref{table:nonlin}. We note that the ratio of diffusion coefficients $\tilde{D}$ in Eq.\ \eqref{lamda} is large, which means that quantities in Eqs.\ \eqref{first} and \eqref{second} take dimensionless values covering several order of magnitudes O$(10^{-2})$ - O$(10^6)$. This is because the nutrient density distribution evolves on much faster time scales than the cells, which creates challenges for the numerical methods that we use below. Since the algorithm must run over a long time, the (nutrient) terms associated with the O$(10^6)$ parameters equilibrate very rapidly by a time $t \sim$ O$(10^{-6})$, compared to the slower (cells evolution) processes which take times $t \sim$ O$(10^{2})$. Consequently, tempering the large valued parameters, say by setting $(10^{6}) \mapsto 1$ for the large parameters, has little effect on the long term results, but greatly helps in the running of the numerical code. We therefore select the parameter set given in Table \ref {table:nonlin} and henceforth use these as our standard parameter set. We also present results below, illustrating how the long time results for $\rho(\textbf{r},t)$ depend only very weakly on the value of $\tilde{D}$, as it is varied in the range $1\leq\tilde{D}\leq10^2$.


\begin {table}[t!]
\caption{Model parameters and their units. Values marked with an asterisk (*) are estimates from the Appendix.\label{table:nonlin2}}
\centering
\begin{tabular}{|c|  c| c| c|}
\hline \hline
Symbol  & typical value & Unit & Source\\ [0.5ex]
\hline
$\rho(\textbf{r},t)$  &$3\times 10^{5} * $ &$ cm^{-2}$ & estimated\\  \hline
$n(\textbf{r},t)$ &3 *& $mg/L$ &estimated \\  \hline
$ V_{int}(r)$&$\varepsilon$&$Joule$& estimated\\  \hline
$N(t)$& $\rho_0L^2$&dimensionless&$\S \ref{nm}$\\  \hline \hline
$R$  &0.001 & $cm$  &\cite{o2010essentials}\\  \hline
$\lambda_m$ &0.00015 *& $L min^{-1} mg^{-1}$ &estimated \\  \hline
$\lambda_d$  &0.00005 *& $min^{-1}$ &estimated\\  \hline
$\lambda_n$ & 3 *&$ min^{-1}$&estimated\\  \hline
$D_c$&$1.3\times10^{-9} *$ & $cm^2 min^{-1}$&estimated\\   \hline
$D_n$& 0.0012&$cm^2 min^{-1}$&\cite{ward1997mathematical}\\  \hline
$\Gamma$&$3 \times 10^{10} $ *& $ min$ $g^{-1}$ &estimated\\  \hline
$T$& 310&$K$&\cite{o2010essentials}\\  \hline
$k_B$ &$1.38 × 10^{-23}$&$Joule/K$&\cite{mohr2012codata}\\  \hline
$\varepsilon$&$\approx k_B T$&$ Joule$&estimated\\  \hline
$\rho_0$&$3 \times 10^5$ *&$cm^{-2}$&$\S \ref{nm}$\\  \hline
$L^2$& $6 \times 10^{-4}$&$cm^2$&$\S \ref{nm}$\\  \hline
$S_n$& 433 *& $mg  L^{-1} min^{-1} cm^{-2}$&estimated\\  \hline
\hline
\end {tabular}
\label{table:nonlin2}
\end{table}
\begin {table}[t!]
\caption{Dimensionless parameter values of the model.
\label{table:nonlin}}
\centering
\begin{tabular}{|c| c| c| c|}
\hline \hline
 Dim.-less param.& Dim. form & Value & Used value \\ [0.7ex]
\hline
$ c_1 $ &$ {R^2 \lambda_d}/{D_c}$ &0.038 &1 \\ \hline
$\tilde D $ &$ {D_n}/{D_c}$ &$10^6$& $1,10,10^2$ \\ \hline
$\tilde S_n$ &${R^2 S_n \lambda_m}/{\lambda_d D_c}$&$10^6$&10,35 \\ \hline
$ \tilde \lambda_n$&${\lambda_n }/{ D_c}$ &$10^6$&1  \\  \hline
$\beta \varepsilon $&$\beta \varepsilon$ &O(1)&1\\[1.0 ex]
\hline
\hline
\end {tabular}
\end{table}

\subsection {Linear stability analysis}\label{LSA}

For $\tilde S_n>0$ and $f(\textbf{r})$=1 there is a unique uniform density steady state that is a stationary solution of Eqs.\ \eqref{first} and \eqref{second}, that is
\begin {equation}\label {perterpation1}
n=n_0=1, ~~~\rho=\rho_0=\tilde S_n/\tilde \lambda_n.
\end {equation}
We now investigate the linear stability of the uniform density state $(\rho_0,n_0)$ to non-uniform perturbations $(\delta \rho(\textbf{r} ,t),\delta n(\textbf{r} ,t))$, with $\lVert \delta \rho \lVert_{\infty}=\xi$ and $\lVert \delta n \lVert_{\infty}= \chi \xi$, where $\xi\ll1$. The analysis also applies more generally to determine the growth or decay of a perturbation about a uniform density state $(\rho_0,n_0)$, with values different to those in Eq.\ \eqref{perterpation1}, \cutout{provided $c_1\ll \xi$,} i.e.\ the timescale for cell repositioning in response to the perturbation is much faster than cell growth\correction{; we note $c_1\ll 1$ from data, see Table  \ref{table:nonlin}}. \cutout{When $c_1>\xi$, the analysis is only relevant for $n_0=1$.} Note that it is the parameter values where the uniform system is unstable (and forms peaks) that are of relevance biologically.

To determine the linear stability of the flat state, we assume that the cell density profile take the form 
\begin {align}\label{perterpation}\nonumber
\rho&=\rho_0+\delta \rho(\textbf{r} ,t)\\
&=\rho_0+\xi e^{i(\textbf{k.r})+ \omega t},
\end {align}
and the nutrient density profile
\begin {align}\label{perterpation2}\nonumber
n&=n_0+ \delta n(\textbf{r} ,t)\\
&=n_0+ \chi \xi e^{i(\textbf{k.r})+ \omega t},
\end {align}
where $0<\xi\ll1$ is the initial amplitude of the sinusoidal perturbation that has wavenumber $k=|\textbf{k}|$, $\chi$ is the ratio between the amplitude of the modulation in the two components, and the growth or decay rate of the perturbations is given by the dispersion relation $\omega=\omega(k)$.  
Substitution of Eqs. \eqref{perterpation} and \eqref{perterpation2} into the dynamic equation \eqref{first}, and then linearising in $ \delta \rho$ we obtain (c.f.\ \cite{ArEv04,archer2014solidification})
\begin{equation}\label{f63}
\omega(k)=-k^2\big[1+\rho_0  \beta \varepsilon \hat V(k)\big]+ c_1 \left(n_0+ \rho_0 \chi -1 \right),
\end{equation}
where $ \hat V(k) $ is the Fourier transform of the pair potential. Since we have assumed the system is in 2D, the Fourier transform is
\begin {equation}\label{FT}
\hat V(k)= {\int d {\mathbf{r}} e^{i\textbf{k.r}} \tilde V_{int}(\mathbf{r})}=2 \pi \int_0^{\infty} r \tilde V_{int}(r)J_0(kr) dr
\end{equation}
where $J_0(x)$ is the Bessel function of order 0.

The limit of linear stability is defined as the locus of points in parameter space where the maximum in the dispersion relation \eqref{f63} is at zero, i.e.\ $\omega(k=k_c)=0$, where $k_c$ is the wave vector where $\omega(k)$ is maximum, where $\frac{d \omega}{dk}|_{k=k_c}=0$. In the case of \cutout{$c_1\ll\xi\ll1$}  \correction{$c_1\ll1$} we have
\begin {equation}
1+\rho_0  \beta \varepsilon \hat V(k=k_c) \approx 0,
\end {equation}
where \correction{$k_c\approx 5.1$} \cutout{$k_c=5.3$} and \correction{$\hat V(k_c)\approx -0.16$} \cutout{$\hat V(k_c)=-0.29$} (recall that in the nondimensionalisation we effectively set the unit of length $R=1$), which implies that the locus of where the system becomes linearly unstable is  
\begin {equation}\label{threshold}
\rho_0\approx \frac{1}{ \beta \varepsilon |\hat V(k_c)|}
\end {equation}
which in the density $\rho_0$ versus ``dimensionless temperature'' $k_BT/\varepsilon=1/\beta \varepsilon$ plane is a straight line passing through the origin \cite{archer2014solidification}. For densities greater than this value, the system is linearly unstable. Note that even though we have assumed \correction{$c_1\ll1$} \cutout{$c_1\ll\xi\ll1$} in the derivation, it turns out that even for \cutout{larger} $c_1$ \correction{$=O(1)$}, Eq.\ \eqref{threshold} gives a good estimate for where the system is linearly unstable. \cutout{This is the regime relevant to modelling the density distribution of the cells.} \correction{Given the data in Table \ref{table:nonlin}, the analysis suggests that dominant terms governing instability is the cell density and the cell-cell interaction parameters; cell growth and nutrient consumption rates are secondary to this process.}

 \subsection {Numerical results for the cell evolution}\label{nm}

The coupled equations \eqref{first} and \eqref{second} are solved numerically using the method of lines. The density profiles are discretised on a spatially uniform grid, with the convolution integral evaluated in Fourier space using fast Fourier transforms, whilst for the time stepping the Adam-Bashforth method is implemented, via the freeware ODEPACK routine LDSODE \cite{hindmarsh1983odepack,hindmarsh2002serial}. We note that this time stepping method is significantly faster than the Euler time stepping routines used for the similar problem in \cite{archer2014solidification}. We note that all quantities shown in the figures, including those of Section \ref{sec:num_res_2}, are dimensionless.\\

\subsubsection {Results with homogeneous nutrient source}
We assume initial conditions  
\begin {align}\label {rand}\nonumber
\rho(\textbf{r},0)=1+\gamma (\textbf{r})\\
n(\textbf{r},0)=1~~~~~~~~~
\end {align}
where $\gamma (\textbf{r})$ is a small amplitude random variable and $\gamma (\textbf{r})\sim U(0,1)$, where $U$ is a uniform distribution. We set the dimensionless model parameters to be $c_1=1$, $\beta \varepsilon=1$, $\tilde \lambda_n=1$ and $\tilde D=1$. We set the area of the domain in which the model is solved to be $25.6 \times 25.6$, with grid spacing $\Delta x=0.1$ (smaller values were also tested, but this value is normally sufficiently small) and periodic boundary conditions on all sides. We set the nutrient source to be uniform $f(\textbf{r})=1$, with amplitude $\tilde S_n=10$.

\begin{figure}[t]
\begin {center}	
	\includegraphics[width=0.48\textwidth]{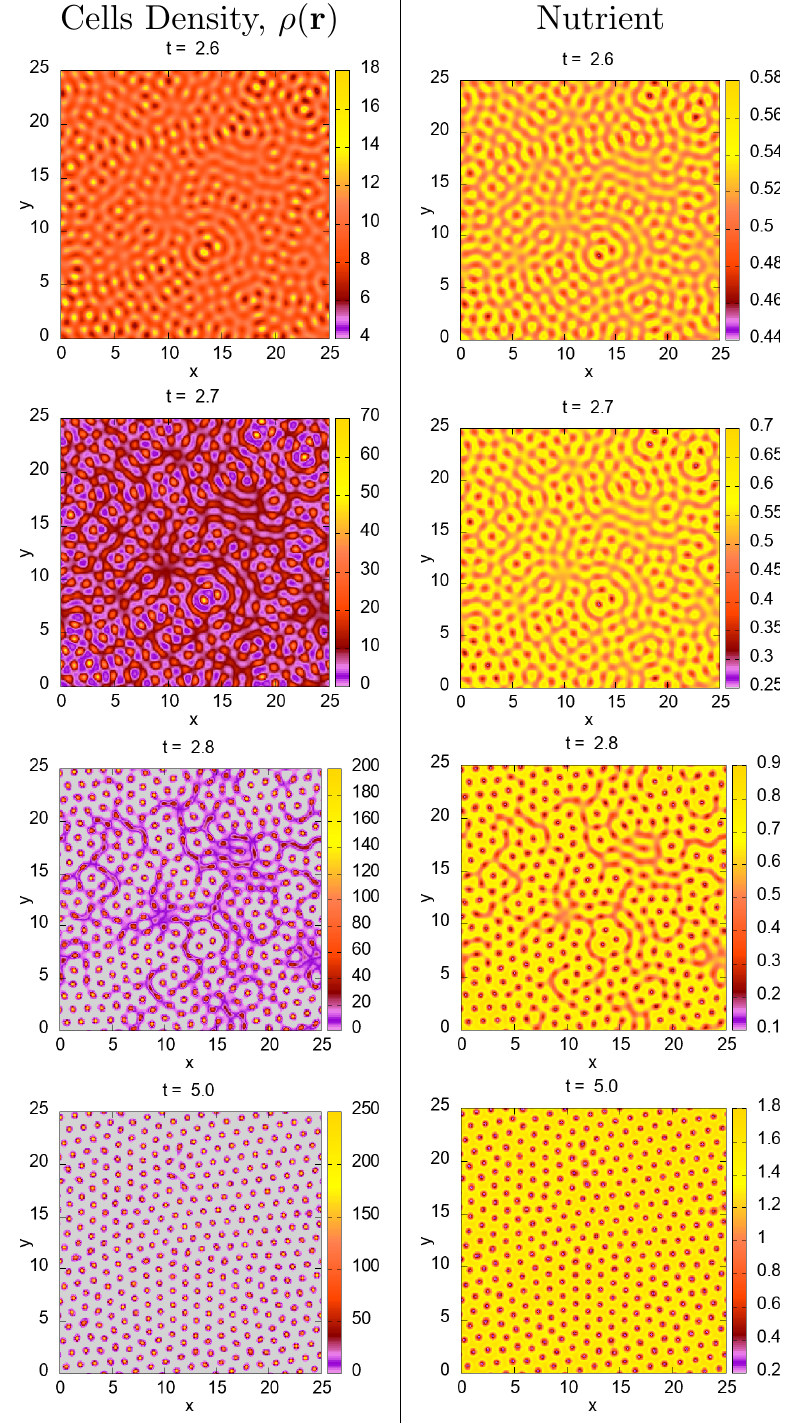}
	\end{center}
	\caption{%
 Density of the cells (left) and local nutrition concentration (right) over time. We assume that the population growth constant $c_1=1$ and the energy scale in the interaction potential between cells $\beta \varepsilon=1$. The diffusion coefficient ratio $\tilde D=1$. The nutrient source is homogeneous with $f(\textbf{r})=1$ and $\tilde S_n=10$, and the nutrient uptake rate $\tilde \lambda_n=1$. The area of the domain is $25.6^2$ and $\Delta x=0.1$.
}
\label{res3}
\end{figure}

\begin{figure}[t]
	\begin {center}
		\includegraphics[width=.48\textwidth]{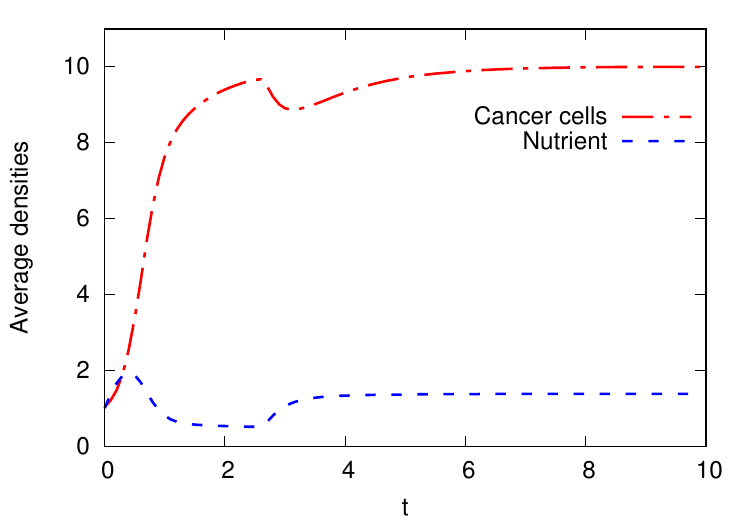}
	\end{center}
	\caption{
	The average cell density, [see Eq.\eqref{acd}] and the average nutrient density [see Eq.\eqref{annd}], corresponding to the results in Fig. \ref{res3}.
	}
	\label{rhogsn}
\end{figure}

In Fig.\ \ref{res3}, the plots in the left hand column are the density profile of the cells at a series of different times ($t$=2.6, 2.7, 2.8 and 5), while the right hand column displays plots of the local nutrient concentration. From the left column, it is clear that the total density of cells increases with time, as can also be seen in Fig.\ \ref{rhogsn} where we plot the average cell density and nutrient density over the whole system as a function of  time, which are defined in Eqs.\ \eqref{acd} and \eqref{annd}. We see the peaks (i.e.\ locations of the centres of the cells) grow and split to fill the entire domain, due the fact that there is a source of nutrient everywhere, in contrast to the behaviour seen for example in Fig. \ref{result1} where the source of nutrient is localised along the mid-line of the system. In Fig.\ \ref{rhogsn} we see that initially the nutrient density increases, due to the low initial average cell density. Then, at $t \approx 0.5$, whilst the cell density increases, the nutrient density starts to decrease, due to the increased consumption. Over the time $2\lesssim t\lesssim 3$ the peaks in the cells density distribution form. Consequently, the nutrient concentration then increases again at $t\approx 3$. After this, $\bar n(t)$ is roughly a constant $\approx 1.2$, as shown in Fig.\ \ref{rhogsn}. The cell density continues to slowly increase to plateau at a constant value $\approx 10$ at the time $t \approx 6$.

 \subsubsection {Results with inhomogeneous nutrient source}

\begin{figure*}[t]
\begin {center}
	\includegraphics[width=.7\textwidth]{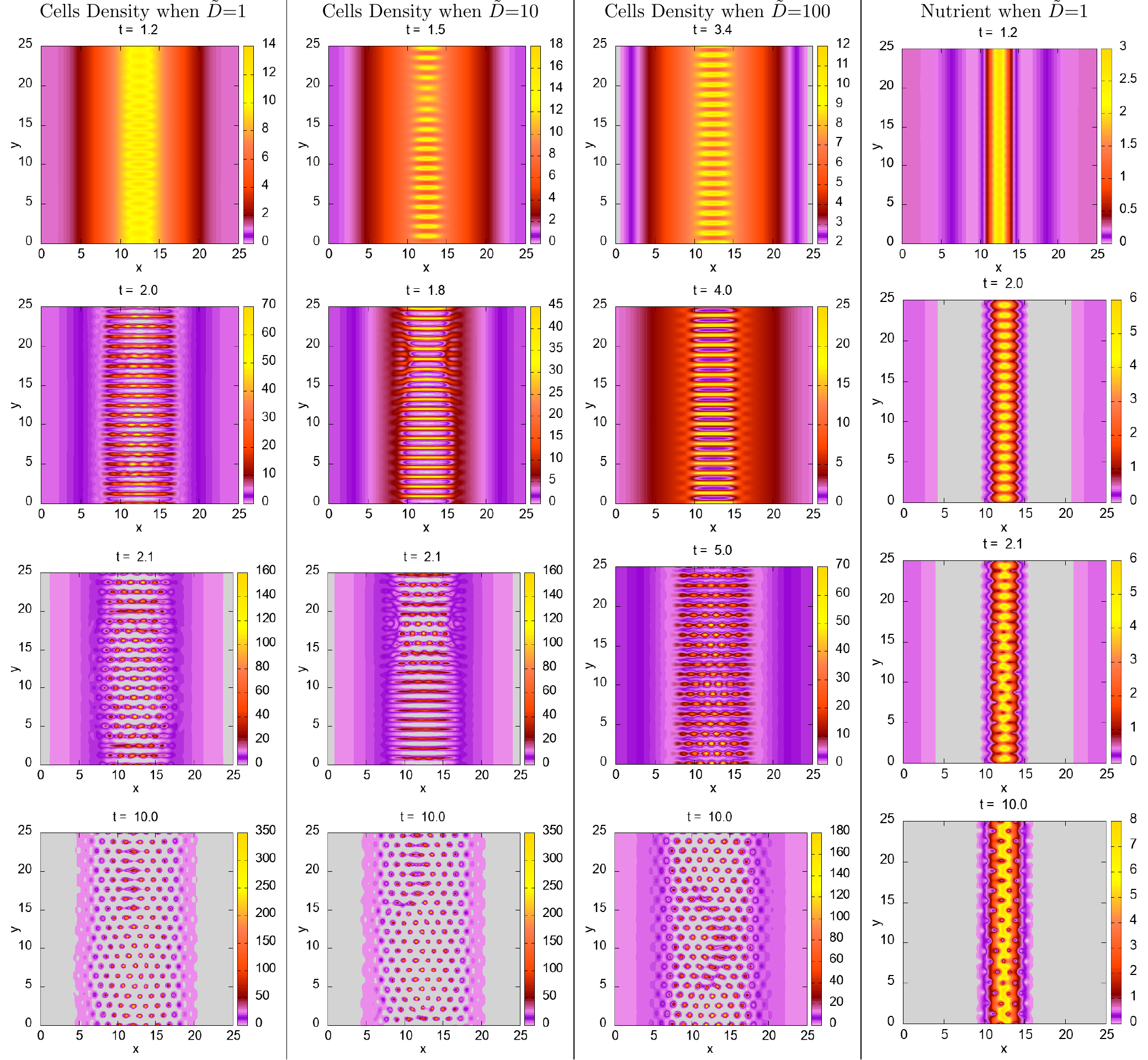}
	\end{center}
\caption{%
The local density of the cells (left three columns, for $\tilde D$=1, 10 and 100, from left to right) and the nutrient density for $\tilde D=1$ (right hand column). The population growth constant $c_1=1$ and the energy scale in the interaction potential between cells $\beta \varepsilon=1$. The nutrient source term has $\tilde S_n=35$ with $f(\textbf{r})$ given in Eq.\eqref{ge} and nutrient uptake rate $\tilde \lambda_n=1$. The area of the system is $25.6^2$, with grid spacing $\Delta x=0.05$.
}
\label{result1}
\end{figure*}

Fig.\ \ref{result1} compares results for the cell density profile time evolution for three different values of $\tilde D=1$, 10 and $100$ (from left to right). For example, the results in the left hand column of Fig.\ \ref{result1} shows the evolution of cell density, displaying snapshots for the times $t$=1.2, 2, 2.1 and 10. In these cases the nutrient source is located along the vertical mid line of the system [c.f.\ Eq.\ \eqref{ge}]. From an initial randomised distribution, the cell density grows in the vicinity of central nutrient source. When the density is sufficiently large, an instability (c.f.\ Sec.\ \ref{LSA}) leads first to a striped pattern and then peaks. The density peaks (i.e.\ cells) are arranged in a roughly hexagonal pattern, which also impacts the nutrient distribution. The right hand column of Fig.\ \ref{result1} show the time evolution of the nutrient density for the case $\tilde D=1$, corresponding to the left hand column cell density profiles.

\begin{figure}[t]
	\begin {center}
		\includegraphics[width=.48\textwidth]{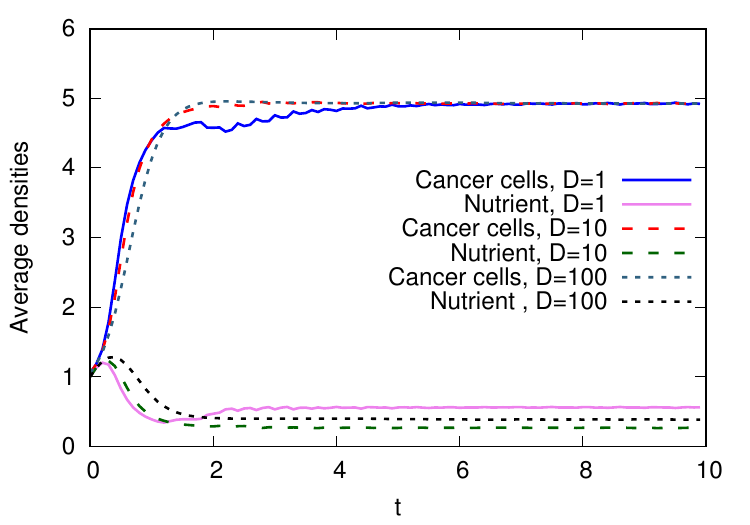}
	\end{center}
	\caption{
	The average cell density, [see Eq.\ \eqref{acd}] and the average nutrient density [see Eq.\ \eqref{annd}], corresponding to the results in Fig.\ \ref{result1} when the diffusion coefficient $\tilde D$=1, 10 and 100 respectively.
	}
	\label{rho100}
\end{figure}

In Fig.\ \ref{rho100} we display plots of the total cell density and nutrient density calculated using Eqs.\ \eqref{acd} and \eqref{annd}, corresponding to Fig.\ \ref{result1}. These results are for three very different values of $\tilde D=1$, 10 and $100$. Nonetheless, we see that in all three cases the results are all qualitatively rather similar, which demonstrates that for $\tilde D\gtrsim1$ the results do not qualitatively depend on the precise value of $\tilde D$. Recall that in Sec.\ \ref{LSA} we note that the true value is $\tilde D\approx 10^6$ [see also the Appendix and Eq.\ \eqref{diffusion}], but also argue that we do not need to have such a large value. Owing to the qualitative similarity of the results shown in Fig. \ref{result1}, we see that smaller values of $\tilde D\approx 10$ are acceptable.

The similarities for different values of the diffusion coefficient ratio $\tilde D$ can also be seen from the results in Fig.\ \ref{rho100}, whereby the steady value of $\bar \rho \approx 5$ and $\bar n \approx 0.5$ is reached by $t \approx 4$. Note that for the smaller $\tilde D=1$ case there are small amplitude oscillations in both the cell and nutrient average densities for $t>2$. These are due to new cells being formed and then dying in a periodic fashion.

\begin{figure}
	\begin {center}
	\includegraphics[width=.5\textwidth]{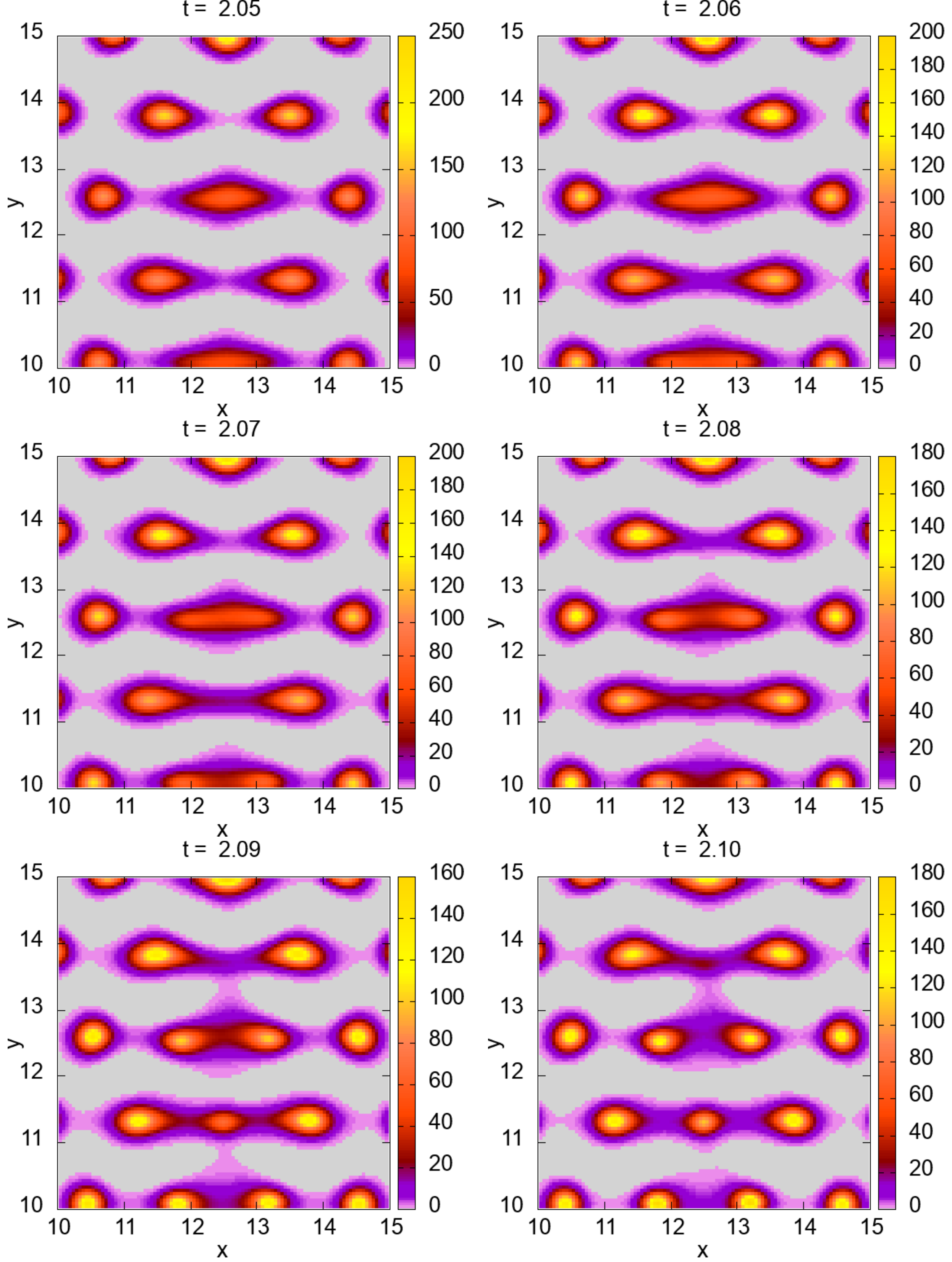}
	\end{center}
 \caption{%
 Snapshots of several peak splitting events that occur between the times $t=2.05$ and $t=2.10$. The figures above are in time increments of  $0.01$ going from top left to bottom right, corresponding to the profiles plotted in the left hand column of Fig.\ \ref {result1}, which are for $\tilde D=1$.}
\label{result23}
\end{figure}

By $t \approx 10$ the cell density profiles in Fig.\ \ref{result1} no longer change qualitatively, however they are {\em not} stationary. We see that around the nutrient source along the line $x=L/2$, we have a region where the peaks grow and then split -- modelling cell division -- and then move away from the nutrient source, where they subsequently die due to the lack of nutrient away from the centre line. In Fig.\ \ref{result23} we display a magnification of the cell density profile to highlight these mitotic events. The sequence of snapshots in Fig. \ref{result23} illustrates the cell splitting events that occurs between the times $t=$2.05 and $t=$2.10 with time increments of  $0.01$. We observe that a peak first elongates and then splits to form new peaks which remarkably mirrors a mitotic event. In the fourth row in Fig.\ \ref{result23}, a peak spontaneously emerges between two existing ones, describing the average location of a new cell resulting from mitosis of one of the cells either side of it.

\section{Competition between cancer and healthy cells}\label{sec61}

In this section we extend the model presented in the previous section to include a second species of cells. Our aim is to study the competition between cancer cells and healthy cells. We denote the density of the cancerous and the healthy cells as $ \rho_{1}$ and $ \rho_{2}$, respectively. The generalisation of Eqs.\ \eqref{vvv} and \eqref{g} is
\begin{eqnarray} \nonumber \label{eq:666}
\frac{\partial \rho_{1} (\textbf{r},t)}{\partial t}&=&\Gamma_{1}\nabla\cdot\left[ \rho_{1}(\textbf{r},t)\nabla\left(\frac{\delta {\cal F}[\rho_{1},\rho_{2}]}{\delta\rho_{1}(\textbf{r},t)}\right)\right] \\
&&+[ \lambda_{m1}n(\textbf{r},t)-\lambda_{d1}] \rho_{1} (\textbf{r},t),
\end{eqnarray}
\begin {eqnarray} \nonumber \label{eq:777}
\frac{\partial \rho_{2} (\textbf{r},t)}{\partial t}&=&\Gamma_{2}\nabla\cdot\left[ \rho_{2}(\textbf{r},t)\nabla\left(\frac{\delta {\cal F}[\rho_{1},\rho_{2}]}{\delta\rho_{2}(\textbf{r},t)}\right)\right] \\
&&+[ \lambda_{m2}n(\textbf{r},t)-\lambda_{d2}] \rho_{2} (\textbf{r},t),
\end{eqnarray}
 \begin{eqnarray}\label{888}
  \frac{\partial n(\textbf{r},t)}{\partial t}&=&D_n \nabla^{2}n(\textbf{r},t)+S_n f(\textbf{r})-\lambda_{n1}\rho_{1} (\textbf{r},t) n(\textbf{r},t)) \nonumber \\
  &&-\lambda_{n2}\rho_{2} (\textbf{r},t) n(\textbf{r},t),
  \end{eqnarray}
where $\lambda_{mi}, \lambda_{di}, \lambda_{ni}$ and $\Gamma_i$ have the same as their counterparts in Section \ref{app} for species $i$. The generalisation of DDFT to describe a two component colloidal suspension was discussed in \cite {archer2005dynamical}. The above reduces to this DDFT if the BD terms are set to zero. 

For such a binary system we may approximate the intrinsic Helmholtz free energy of the system as in \cite{archer2005dynamical,likos2001effective}, namely
  \begin{eqnarray} \nonumber
{\cal F}[\{\rho_{i}(\textbf{r},t)\}]=k_B T\sum_{i=1}^{2} \int d\textbf{r} \rho_{i}(\textbf{r},t)\left( \ln [\Lambda_{i}^d\rho_{i}(\textbf{r},t)]-1\right)\\
+\frac{1}{2}\sum_{i,j=1}^{2}\int d\textbf{r}\int d\textbf{r}{'} \rho_{i}(\textbf{r},t)\rho_{j}(\textbf{r}{'},t)V_{ij}(|\textbf{r}-\textbf{r}{'}|),~~~~~ \label{eq68}
\end{eqnarray}
where $V_{ij}$ are the pair interactions potentials, discussed further below. The indices $i,j=1,2$ label the two different species of particles (healthy and cancer); we assign 1 for cancer cells and 2 for healthy cells. Substituting Eq.\ \eqref{eq68} into Eqs.\ \eqref{eq:666} and \eqref{eq:777}, we obtain
\begin{eqnarray}\nonumber \label{6666}
\frac{\partial \rho_{1}(\textbf{r},t)}{\partial t}&=& \nabla \cdot \bigg [ \Gamma_{1} \rho_{1}(\textbf{r},t)\nabla \bigg(k_B T \ln (\Lambda_1^d \rho_{1}({\mathbf{r}},t) \\
&&  + \int d\textbf{r}{'}\rho_{1}(\textbf{r}{'},t)V_{11}(|\textbf{r}-\textbf{r}{'}|) \nonumber\\
&&+\int d\textbf{r}{'}\rho_{2}(\textbf{r}{'},t)V_{12}(|\textbf{r}-\textbf{r}{'}|)\bigg )\bigg] \nonumber\\
&& +\big[ \lambda_{m1}n(\textbf{r},t)-\lambda_{d1}\big] \rho_{1} (\textbf{r},t)
\end{eqnarray}
and
\begin{eqnarray} \nonumber \label{7777}{}
\frac{\partial \rho_{2}(\textbf{r},t)}{\partial t}&=& \nabla \cdot \bigg [ \Gamma_{2} \rho_{2}(\textbf{r},t)\nabla \bigg(k_B T \ln (\Lambda_2^d \rho_{2}({\mathbf{r}},t))   \nonumber \\
&&+  \int d\textbf{r}{'}\rho_{1}(\textbf{r}{'},t)V_{21}(|\textbf{r}-\textbf{r}{'}|) \nonumber \\
&&+\int d\textbf{r}{'}\rho_{2}(\textbf{r}{'},t)V_{22}(|\textbf{r}-\textbf{r}{'}|)\bigg )\bigg]  \nonumber \\
&& +\big[ \lambda_{m2}n(\textbf{r},t)-\lambda_{d2}\big] \rho_{2} (\textbf{r},t),
\end{eqnarray}
where $\Lambda_i$ are the thermal de Broglie wavelengths for species $i$. As in Sec.\ \ref{sec1} we model the cell-cell interactions via soft, purely repulsive and radially symmetric pair potentials given by
\begin{equation}\label{dafra}
V_{ij}(r)= \varepsilon_{ij} e^{-(r/R_{ij})^{4}},
\end{equation}
where the parameters $\varepsilon_{ij}$ specify the strength of the repulsion between pairs of cells of species $i$ and $j$ and $R_{ij}$ define the range of the interactions. Thus, we choose $R_{11} \ge R_{22}$, since cancer cells are generally slightly larger than healthy cells and we choose $\varepsilon_{12} >  \varepsilon_{11}=\varepsilon_{22}$, so that peaks of the different species do not occur at the same point in space. In some cases we choose $R_{12} =\frac{1}{2} (R_{11}+R_{22})$, but we also consider cases where $R_{12} > \frac{1}{2} (R_{11}+R_{22})$ since this promotes demixing of the two cell species and also $R_{12} < \frac{1}{2} (R_{11}+R_{22})$ which promotes penetration of the cancer cells in between the healthy cells \cite{archer2005dynamical,likos2001effective,archer2001binary}.

\subsection{Nondimensionalisation}\label{nond}
We nondimensionlise the system of integro-partial differential equations given in Eqs. \eqref{6666}, \eqref{7777} and \eqref{888} in a manner similar to previously, using
$ t= \frac{R_{11}^{2} t^{*}}{D_c}$,
$x=\frac{x^{*}}{R_{11}}$,
$y=\frac{y^{*}}{R_{11}}$,  
$ \rho_{1}=\frac{\rho_{1}^{*}}{R_{11}^{2}}$,
$ \rho_{2}=\frac{\rho_{2}^{*}}{R_{11}^{2}}$,
$ n= \frac{\lambda_{d1} n^{*}}{\lambda_{m1}}$ and 
$V_{ij}(r/R_{11})= \varepsilon_{ij}\tilde {V}_{ij}(r^*)$,
where the asterisked quantities are dimensionless and $D_c = \Gamma_1 k_B T$. Here, the scaling on space is based on the range of the interaction between two cancer cells, $R_{11}$. Defining the dimensionless parameters [c.f.\ Eq.\ \eqref{lamda}]
{\allowdisplaybreaks \begin{eqnarray} \nonumber
c_1&=&\frac {R_{11}^2 \lambda_{d1}}{D_{c}}, ~ ~ ~ c_2=\frac {R_{11}^2 \lambda_{m2}\lambda_{d1}}{D_{c}\lambda_{m1}}, ~\alpha=\frac { \lambda_{d2}\lambda_{m1}}{\lambda_{d1}\lambda_{m2}},
\\
\tilde D_2 &=&\frac {\Gamma_2}{\Gamma_1}, ~ ~ ~\tilde D_n =\frac {D_n}{D_{c}}, ~ ~ ~ \tilde S_n=\frac{R_{11}^2 S_n \lambda_{m1}}{\lambda_{d1} D_{c}},
\nonumber \\
\tilde \lambda_{n1}&=&\frac{\lambda_{n1}}{{D_{c}}}, ~~ \tilde \lambda_{n2}=\frac{\lambda_{n2}}{ {D_{c}}},\nonumber
\end{eqnarray}
noting that $\tilde D_2$ is the ratio of the diffusion coefficients of healthy cells to cancer cells. We get 
\begin{eqnarray}\nonumber \label{koo}
\frac{\partial \rho_{1}(\textbf{r},t) }{\partial t }&=&\nabla^2 \rho_{1}(\textbf{r},t)\nonumber \\
&&+\nabla \cdot \left ( \rho_{1} (\textbf{r},t) \nabla \int d \textbf{r}{'}\rho_{1}(\textbf{r}{'},t)  \beta \varepsilon_{11} \tilde V_{11}(|\mathbf{r-r}'|)\right ) \nonumber\\
&&+\nabla \cdot \left ( \rho_{1} (\textbf{r},t) \nabla \int d \textbf{r}{'}\rho_{2}(\textbf{r}{'},t)  \beta \varepsilon_{12} \tilde V_{12}(|\mathbf{r-r}'|)\right ) \nonumber \\
&&+c_1 \big[ n(\textbf{r} ,t)-1 \big] \rho_{1} (\textbf{r},t),
\end{eqnarray}
\begin{eqnarray} \label{ko}
\frac{\partial \rho_{2}(\textbf{r},t) }{\partial t }&=& \tilde D_2 \nabla^2 \rho_{2}(\textbf{r},t)  \nonumber \\
&&+ \nabla \cdot \left ( \rho_{2} (\textbf{r},t) \nabla \int d \textbf{r}{'}\rho_{1}(\textbf{r}{'},t)  \beta \varepsilon_{21} \tilde V_{21}(|\mathbf{r-r}'|)\right ) \nonumber \\
&&+ \nabla \cdot \left ( \rho_{2} (\textbf{r},t) \nabla \int d \textbf{r}{'}\rho_{2}(\textbf{r}{'},t)  \beta \varepsilon_{22} \tilde V_{22}(|\mathbf{r-r}'|)\right )  \nonumber \\ 
&&+c_2 \big[  n(\textbf{r} ,t)-\alpha \big] \rho_{2} (\textbf{r},t),
\end{eqnarray}
\begin{eqnarray}\label{k}
\frac{\partial n(\textbf{r},t)}{\partial t }&=&\tilde D_n \nabla^{2}n(\textbf{r},t)+\tilde S_n f(\textbf{r}) \nonumber \\
&& - \tilde \lambda_{n1} \rho_{1} (\textbf{r},t)  n(\textbf{r},t)- \tilde \lambda_{n2} \rho_{2} (\textbf{r},t)  n(\textbf{r},t).~~~~~
\end{eqnarray}}
Where the asterisks have been dropped for clarity.

\subsection {Parameters values} \label{pv6}
For both the healthy and the cancer cell growth rate parameters, diffusion coefficients and the parameters relating to the nutrient dynamics we use the same values that are argued for in the Appendix. The main change is to make the growth rate parameters for the cancer cells larger than those of the healthy cells in order for them to reproduce and grow faster (or die slower) than the healthy cells. The parameter values are summarised in Table \ref{xxx} and the corresponding dimensionless parameter values are given in Table \ref{table:nonlin6}. The other main addition to the model for both healthy and cancer cells that must be considered are the parameter values in the interaction potential between the different types of cells, given in Eq. \eqref{dafra}. The parameter values we choose are given in Table \ref{xxx}. These values are chosen in order to (i) make the cancer cells either the same size or slightly larger than the healthy cells \cite{derenzini1998nucleolar} and (ii) to make sure the cancer cells do not overlap with the healthy cells.

\begin {table}[t]
\caption{Model parameters and their units. Values marked with asterisk (*) are estimates from Secs.\ \ref{pv} and \ref{pv6}.\label{xxx} }
\centering
\begin{tabular}{|c|  c| c| c|}
\hline \hline
Symbol  & typical value & Unit & Source\\ [0.5ex]
\hline
$\rho_{1}(\textbf{r},t)$  &$3\times 10^{5} *$& $cm^{-2}$ & estimated\\ \hline
$\rho_{2}(\textbf{r},t)$  &$3\times 10^{5} *$& $cm^{-2}$ & estimated\\ \hline
$n(\textbf{r},t)$  &3 *& $mg/L$ &estimated \\ \hline
$V_{11}(r)$&$\varepsilon_{11}$&$Joule$& estimated\\ \hline
 $V_{12}(r)$&$\varepsilon_{12}$&$Joule$& estimated\\ \hline
 $V_{22}(r)$ &$\varepsilon_{22}$&$Joule$& estimated\\ \hline \hline
$R_{11}$  &0.001 & $cm$  &\cite {o2010essentials}\\ \hline
$R_{22}$ &0.0009 & $cm$  &\cite {o2010essentials}\\  \hline
$\lambda_{m1}$ &0.00015 *& $L min^{-1} mg^{-1}$ &estimated \\ \hline
$\lambda_{m2}$ &0.000015 *& $L min^{-1} mg^{-1}$ &estimated \\ \hline
$\lambda_{d1}$  &0.00005 *& $min^{-1}$ &estimated\\ \hline
$\lambda_{d2}$  &0.000005 *& $min^{-1}$ &estimated\\ \hline
$\lambda_{n1}$& 3 *&$min^{-1}$&estimated\\ \hline
$\lambda_{n2}$& 3 *&$min^{-1}$&estimated\\ \hline
$D_{c}$&$1.3\times10^{-9}$ *& $cm^2$ min$^{-1}$&estimated\\  \hline
$D_{h}$&$1.1\times10^{-9}$ *& $cm^2$ min$^{-1}$&estimated\\ \hline
$D_n$ & 0.0012&$cm^2 min^{-1}$&\cite {ward1997mathematical}\\ \hline
$\Gamma_{1}$ &$3 \times 10^{10} $&$min ~g^{-1}$ &$\S \ref{nond}$\\ \hline
$\Gamma_{2}$ &$2.5 \times 10^{10}$  &$min ~g^{-1}$ &$\S \ref{nond}$\\ \hline
$T$&310 &$K$&\cite {o2010essentials}\\ \hline 
 $k_B$ &$1.38 × 10^{-23}$&$Joule/K$&\cite{mohr2012codata}\\ \hline
 $\varepsilon_{11}$&$1 k_B T$& $Joule$& estimated\\ \hline
 $\varepsilon_{12}$&$1.5 k_B T$& $Joule$& estimated\\ \hline
 $\varepsilon_{22}$&$1 k_B T$& $Joule$& estimated\\ \hline
 $\rho_0$ &$3 \times 10^5$  &cm$^{-2}$&$Table ~\ref{table:nonlin2}$\\ \hline
 $L^2$& $6 \times 10^{-4}$&$cm^2$&$Table ~\ref{table:nonlin2}$\\ \hline
$S_n$ & 433 *&$ mg  L^{-1} min^{-1} cm^{-2}$&estimated\\ \hline
\hline
\end {tabular}
\end{table}
\begin {table}[t]
\caption{Dimensionless parameter values of the model. $\gamma(\textbf{r})$ is given in Eq. \eqref{rand}.\label{table:nonlin6}}
\centering
\begin{tabular}{|c| c| c| c| }
\hline \hline
Nondim.p & Dim. form &  value& Used value  \\ [0.7ex]
\hline
$\rho_{1}^{*}$& ${ \rho_{1}}/{ \hat \rho_{1}}$ &  1&$6+\gamma(\textbf{r})^*$  \\ \hline
$\rho_{2}^{*}$& ${ \rho_{2}}/{ \hat \rho_{2}}$ &  1&$6+\gamma(\textbf{r})^*$  \\ \hline
$n^{*} $& $ {n}/{\hat n}$& 3&3 \\ \hline
$ c_1 $ & $ {R_{11}^2 \lambda_{d1}}/{D_{c}}$ &0.038 &0.5, 0.6 \\ \hline
$ c_2 $ & $\ {R_{11}^2 \lambda_{m2}\lambda_{d1}}/{D_{c}\lambda_{m1}}$ &0.0038 &0.5, 0.6 \\ \hline
$  \alpha $ & $ { \lambda_{d2}\lambda_{m1}}/{\lambda_{d1}\lambda_{m2}}$ & 1 & 2 \\ \hline
$\tilde D_2 $ & $ {D_h}/{D_{c}}$ &$1.1$&1  \\ \hline
$\tilde D_n $ & $ {D_n}/{D_{c}}$ &$10^6$&1  \\ \hline
$\tilde S_n$ & ${R_{11}^2 S_n \lambda_{m1}}/{\lambda_{d1} D_{c}}$&$10^6$&8, 9 \\ \hline
$ \tilde \lambda_{n1}$& ${\lambda_{n1} }/{ D_{c}}$ &$10^6$& 1 \\  \hline
$ \tilde \lambda_{n2}$& ${\lambda_{n2} }/{ D_{c}}$ &$10^6$ & 1\\  \hline
$ \varepsilon_{11} \tilde {V}_{11}(r^*)$& $V_{11}(r/R_{11})$ &See Eq. \eqref{dafra}&-\\ \hline
$ \varepsilon_{12} \tilde {V}_{12}(r^*)$& $V_{12}(r/R_{11})$ &See Eq. \eqref{dafra}&-\\ \hline
$ \varepsilon_{22} \tilde {V}_{22}(r^*)$& $V_{22}(r/R_{11})$ &See Eq. \eqref{dafra}&-\\ [1.0 ex]
\hline
\hline
\end {tabular}
\end{table}

\subsection{Linear stability analysis for two species model}\label{sec:LSA_2}

The governing equations for the time evolution of the density profile of the cancer cells, the healthy cells and the nutrient are given by Eqs.\ \eqref{koo}--\eqref{k}. We note for $\alpha \neq $1 there is no spatially uniform positive steady-state to this system. We consider here the linear stability of uniform state $\rho_1= \rho_1^b>0$ and $\rho_2=\rho_2^b>0$ for the case $c_1,c_2 \ll \xi\ll1$, where $\xi$ is the amplitude of the density perturbation\correction{; the small magnitude of  $c_1$ and $c_2$ in comparison to the other parameters is evident from Table \ref{table:nonlin6}}. In setting $c_1=c_2=0$ for the purposes of the linear stability analysis, we are assuming the growth of cells occurs on a much longer time scale than that of the cell motion. This assumption means that the nutrient equation \eqref{k} decouples from Eqs.\ \eqref{koo} and \eqref{ko}\correction{, so that in what follows, stability of a uniform state is predominantly governed by cell density and the cell-cell interaction process}.   

We assume the cell density perturbations are of the form
\begin {align}\label {perterpation3}\nonumber
\rho_1(\textbf{r} ,t)&=\rho_1^b+\delta \rho(\textbf{r} ,t)\\
&=\rho_1^b+\xi e^{i(\textbf{k.r})+ \omega t},
\end {align}
and 
\begin {align}\label {perterpation4}\nonumber
\rho_2(\textbf{r} ,t)&=\rho_2^b+ \chi \delta \rho (\textbf{r} ,t)\\
&=\rho_2^b+ \chi \xi e^{i(\textbf{k.r})+ \omega t},
\end {align}
where $0<\xi\ll1$, $k$ is the wavenumber, $\chi$ is the ratio between the amplitude of the modulation in the two components and the growth or decay rate is determined by the dispersion relation $\omega=\omega(k)$, where $k=|\textbf{k}|$. Substituting Eqs. \eqref{perterpation3} and \eqref{perterpation4} into Eqs. \eqref{koo} and \eqref{ko}, on linearising in $\xi$ we obtain \cite{archer2014solidification}
\begin{equation}\label{mat1}
\omega(k)
\begin{pmatrix} 1 \\ \chi \end{pmatrix}
=\bf{M}
\begin{pmatrix} 1 \\ \chi \end{pmatrix},
\end {equation}
where the matrix
\begin{equation}\label{mat2}
{\bf{M}}=-k^2
\begin{pmatrix} 1+\rho_1^b  \beta \varepsilon_{11} \hat V_{11}(k) & \rho_2^b \beta \varepsilon_{12} \hat V_{12}(k) \\ \rho_1^b  \beta \varepsilon_{21} \hat V_{21}(k) & 1+\rho_2^b \beta \varepsilon_{22} \hat V_{22}(k) \end{pmatrix}.
\end{equation}
We can rewrite the matrix ${\bf{M}}$ as a product of two matrices ${\bf{M}}={\bf{N}} \cdot {\bf{E}}$, where 
\begin{equation}\label{mat3}
{\bf{N}}=
\begin{pmatrix} - \rho_1^b k^2 & 0 \\ 0 & - \rho_2^b k^2 \end{pmatrix},
\end{equation}
and
 \begin{equation}\label{mat4}
{\bf{E}}=
\begin{pmatrix} \big[\frac{1}{\rho_1^b}+  \beta \varepsilon_{11} \hat V_{11}(k)\big] & \beta \varepsilon_{12} \hat V_{12}(k) \\  \beta \varepsilon_{21} \hat V_{21}(k) & \big[\frac{1}{\rho_2^b}+ \beta \varepsilon_{22} \hat V_{22}(k)\big] \end{pmatrix}.
\end{equation}
We can now determine the dispersion relation $\omega(k)$ by calculating the eigenvalues of ${\bf{N}} \cdot {\bf{E}}$,
\begin {equation}
\omega(k)= \frac{\text {Tr}(\bf{{\bf{N}} \cdot {\bf{E}}})}{2} \pm \sqrt{ \frac{\text {Tr}({\bf{{\bf{N}} \cdot {\bf{E}}}})^2}{4}-|\bf{{\bf{N}} \cdot {\bf{E}}}|},
\end {equation}
where $|\bf{{\bf{N}} \cdot {\bf{E}}}|$ denotes the determinant of the matrix $\bf{{\bf{N}} \cdot {\bf{E}}}$ \cite{archer2014solidification}.
When $\omega(k)<0$ for all wave numbers $k$, the system is linearly stable. If, however,  $\omega(k)>0$ for any wave number $k$, then the uniform density state is linearly unstable. Since ${\bf{N}}$ is a (negative definite) diagonal matrix its inverse ${\bf{N}}^{-1}$ exists for all nonzero densities and temperatures, enabling us to write Eq. \eqref{mat1} as the generalised eigenvalue problem
\begin {equation}
({\bf{E}}-{\bf{N}}^{-1} \omega) \hat \chi=0,
\end{equation}
where $ \hat \chi=(1,  \chi)$.
As {\bf{E}} is a symmetric matrix, all eigenvalues are real.\ It follows that the linear stability threshold is determined by $|{\bf{E}}|=0$, i.e. by the condition 
\begin {eqnarray}\label{dk}
D(k) &\equiv& [1+\rho_1^b  \beta \varepsilon_{11} \hat V_{11}(k)][1+\rho_2^b \beta \varepsilon_{22} \hat V_{22}(k) ] \nonumber \\
&&-\rho_1^b\rho_2^b \beta^2 \varepsilon_{12}^2 \hat V_{12}^2(k)=0.
\end{eqnarray}
\begin{figure}
	\begin {center}
	\includegraphics[width=.5\textwidth]{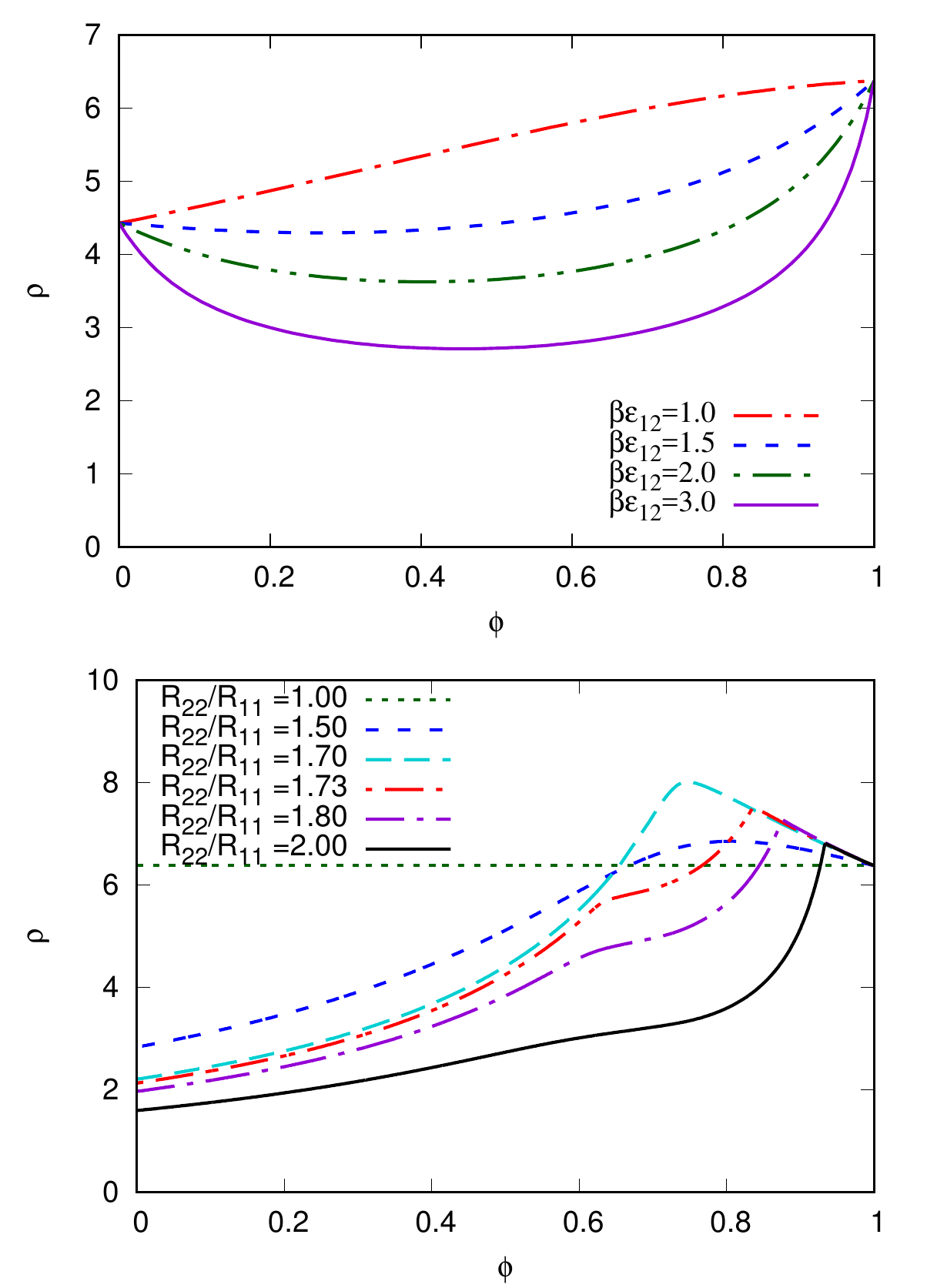}
	\end{center}
	\caption{
	The linear stability threshold for the two species cells [see Eqs.\ \eqref{koo} and \eqref{ko}] plotted in the total density $\rho \equiv \rho_1^b+\rho_2^b$ versus concentration $\phi \equiv \rho_1^b/\rho$ plane. The uniform density state is linearly unstable above this line. The top plot shows the curves for $R_{11}=1$, $R_{22}=1.2$, $R_{12}=1.1$, $ \beta \varepsilon_{11}=\beta \varepsilon_{22}=1$ and for varying $ \beta \varepsilon_{12}$, as given in the key. The curves in the lower plot are for varying $R_{22}=1$, 1.5, 1.7, 1.73, 1.8 and 2. We set the cross-interaction radius $R_{12}=\frac{1}{2}(R_{11}+R_{22})$ and $\beta \varepsilon_{11}= \beta \varepsilon_{12}=\beta \varepsilon_{22}=1$.
	}
	\label{stability2}
\end{figure}
\!\!In Fig.\ \ref{stability2} we display the linear stability threshold for different values of the concentration $\phi \equiv \rho_1^b/\rho$, where $\rho \equiv \rho_1^b+\rho_2^b$ is the total density and $\rho_1^b$, $\rho_2^b$ are the densities of cancer and healthy cells, respectively. For state points above the linear stability threshold lines in Fig. \ref{stability2} the system forms peaks, modelling the distribution of the cells. The instability line is obtained by tracing the locus defined by $D(k_c)=0$ and $D'(k_c)=0$, where $D(k)$ is given in Eq.\ \eqref{dk} and $k_c \neq 0$ is the wave number at the minimum of $D(k)$ [i.e.\ $D(k=k_c)=0$].
Note that as the cell radii ratio $R_{22}/R_{11}$ is increased, the two wavenumbers at which the system can become linearly unstable, $k_c \approx 2 \pi/R_{11}$ or $k_c \approx 2 \pi/R_{22}$, move apart leading to the linear stability threshold developing a cusp, as shown by the ``corners'' in some of the curves in the lower figure of Fig.\ \ref{stability2}. The cusp appears when the two minima in $D(k)$ both satisfy $D(k_c)=0$, and can be determined by simultaneously solving the system of algebraic equations$D(k_c)=D'(k_c)=D''(k_c)=D'''(k_c)=0$. We find that the cusp appears at  $R_{22}/R_{11}$=1.73, $\rho$=8.26 and $\phi$=0.74, 
(red curve in the bottom plot) and is present for $R_{22}/R_{11}>1.73$.

\subsection{Numerical results} \label{sec:num_res_2}
In this section we discuss some representative results showing the competition between healthy and cancer cells, obtained by solving numerically the system of integro-partial differential Eqs.\ \eqref{koo}--\eqref{k} using the numerical methods discussed in Sec.\ \ref{nm}. We investigate the evolution of the cells starting from various different initial arrangements and the effect of the cross-species interaction range $R_{12}$.

\subsubsection {Spread from a few cancer cells within healthy tissue}

In order to model the growth and spread of a tumour within healthy tissue we consider a case where we first initiate the system with one half containing predominantly health tissue, the other half containing cancerous tissue (with uniform densities in each half) and a uniform nutrient density. As the system evolves, peaks form in the two cell density profiles and over time the cancer cells displace the healthy cells till the total average density of healthy cells is small. We then stop the simulation and {\em swap} the labels on the two density profiles, so that the (more realistic) initial condition for the following simulation consists of an array of peaks (cells) in the healthy cell density profile and a low density of cancer cells; i.e.\ for the initial conditions we define $\rho_1(\textbf{r} ,t=0)= \rho_2^\ddagger(\textbf{r} ,t=20)$ and $\rho_2(\textbf{r} ,t=0)= \rho_1^\ddagger(\textbf{r} ,t=20)$, where $\rho_1^\ddagger(\textbf{r} ,t=20)$ and $\rho_2^\ddagger(\textbf{r} ,t=20)$ are the final profiles at time $t=20$ from the preliminary simulation.

\begin{figure}[t]
	\begin {center}
	\includegraphics[width=.45\textwidth]{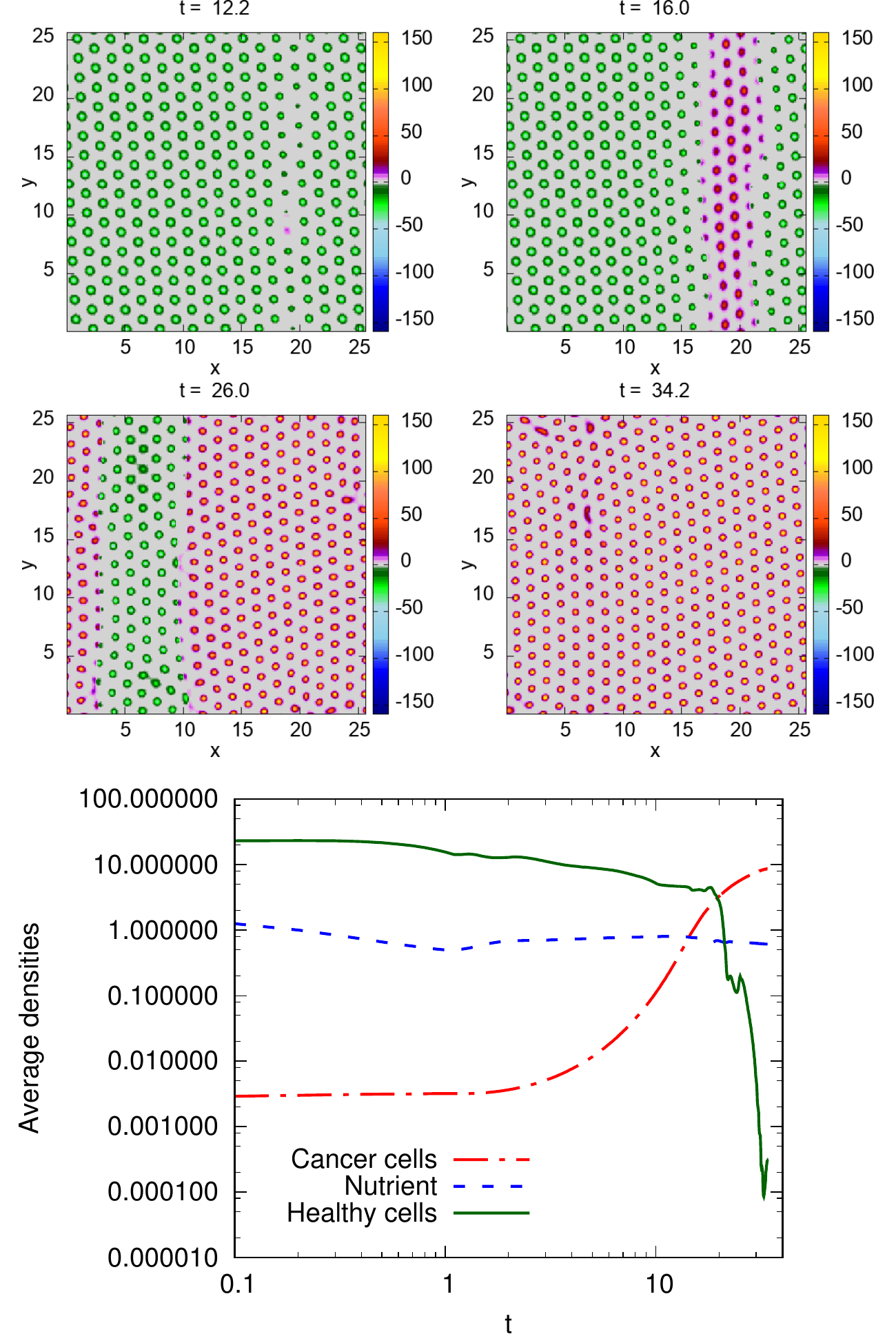}
	\end{center}
	\caption{
	Top four panels: plots of $(\rho_1-\rho_2)$, the density profile of the cancer cells minus the density of the healthy cells, at times $t=0.1$, 16, 26 and 34.2. The nutrient uptake rates $\tilde \lambda_{n1}$=1 and $\tilde \lambda_{n2}$=1, the population growth constants $c_1=c_2=0.5$ and the threshold nutrient concentration for healthy cells $\alpha=2$. The nutrient source is homogeneous, with $f(\textbf{r})=1$ and $\tilde S_n=9$. The area of the domain is $25.6 \times 25.6$ and $\Delta x=\Delta y=0.1$. The cell-cell pair interaction potential parameters are $\beta \varepsilon_{11}$=1, $\beta \varepsilon_{12}$=1.5, $\beta \varepsilon_{11}$=1, $R_{11}=R_{22}=1$ and $R_{12}=0.9$. Bottom: the corresponding average cell density, [see Eq.\ \eqref{acd}] and the average nutrient density [see Eq.\ \eqref{annd}].
	}
	\label{76}
\end{figure}

Snapshots from the subsequent evolution are displayed in Fig.\ \ref{76}. These results are for the population growth constants $c_1=c_2=0.5$ and the threshold nutrient concentration for healthy cells $\alpha=2$. We fix the various cell-cell interaction parameters to be $\beta \varepsilon_{11}=\beta \varepsilon_{22}=1$, $\beta \varepsilon_{12}=1.5$ (so that density peaks of the two different cell types do not overlap), $R_{11}=R_{22}=1$ and  $R_{12}=0.9$. The nutrient uptake rate for cancer cells $\tilde \lambda_{n1}=1$ and for healthy cells $\tilde \lambda_{n2}=1$. The area of the domain in which the model is solved is $25.6 \times 25.6$ and the nutrient source is uniform, with $f(\textbf{r})=1$ and  $\tilde S_n=9$. The diffusion coefficients for both cell species are equal, $\tilde D_{c}=\tilde D_{h}=1$.

In Fig.\ \ref{76} we plot the difference between the density profiles, $(\rho_1-\rho_2)$. Positive values in this quantity correspond to regions where the cancer cells are present (where the peaks are purple-red, with yellow maxima) and negative values where the healthy cells are present (where the peaks are green). In regions that are grey, both densities are low. The Fig.\ \ref{76} profiles are snapshots at the times $t$=12.2, 16, 26 and 34.2. At $t$=12.2 the first cancer cell becomes visible. As time increases, the cancer cells proliferate to form a vertical strip of cancerous tissue, shown in the top right pannel. The fact that it is a vertical strip is due to the original initial conditions. By the time $t=26$ the cancer cells have invaded two thirds of the healthy area and by $t=34.2$ they cover the entire domain, having displaced all the healthy cells.

In the bottom panel of Fig.\ \ref{76}, we plot the average densities of the two species of cells and also of the nutrients, calculated using the two component generalisation of Eq.\ \eqref{acd} and Eq.\ \eqref{annd}, respectively. We see that over time the average nutrient density is roughly constant, but the density of the healthy cells decreases over time, whilst the average density of the cancer cells increases. Interestingly, the average density of the healthy cells does not decrease monotonically; there are instances where there are brief increases, where healthy cells momentarily find gaps around the evolving cancer into which they try and grow. However, the overall trend is for the healthy cells to be displaced and die out.


\subsubsection {Growth of a cancer that is initially small and circular}

\begin{figure}[t]
	\begin {center}
		\includegraphics[width=.5\textwidth]{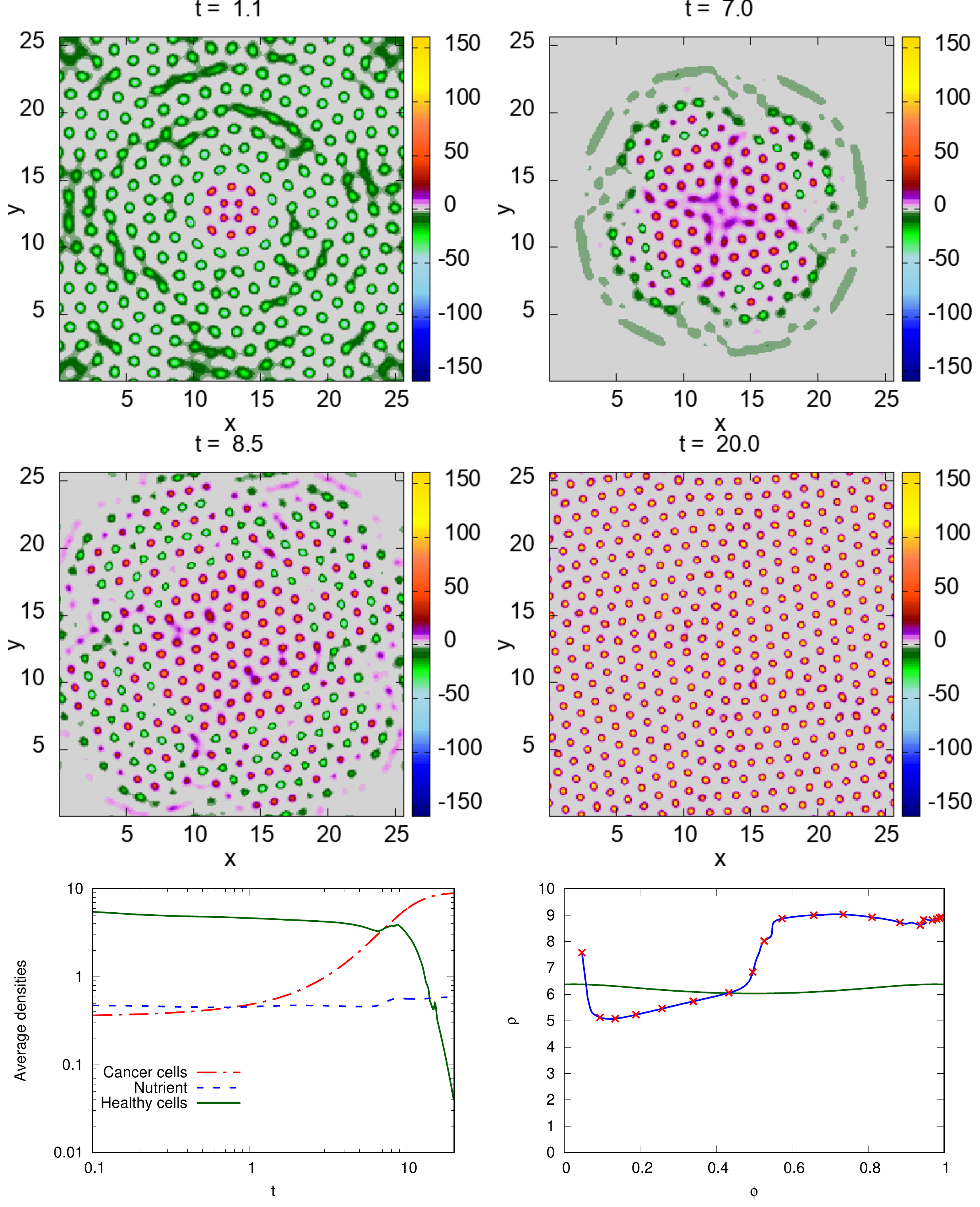}
	\end{center}
	\caption{
	Snapshots of $(\rho_1-\rho_2)$, the density profile of the cancer cells minus that of the healthy cells, at the times $t=0.1$, 6.5, 10 and 20 evolving from the initial conditions defined in Eqs.\ \eqref{IC3} and  \eqref{IC4}. The system parameters are $\tilde \lambda_{n1}=\tilde \lambda_{n2}=1$, $\tilde D_c=\tilde D_2=1$, $c_1=c_2=0.5$ and $\alpha=2$. The nutrient source is homogeneous with $f(\textbf{r})=1$ and $\tilde S_n=9$. The area of the domain is $25.6 \times 25.6$ and $\Delta x=\Delta y=0.1$. The parameters in the pair interaction potentials between the cells are $\beta \varepsilon_{11}=1$, $\beta \varepsilon_{12}=1.5$, $\beta \varepsilon_{11}=1$, $R_{11}=R_{22}=1$ and $R_{12}=0.9$. In the bottom left panel are plotted the corresponding average cell densities [see Eq.\ \eqref{acd}] and the average nutrient density [see Eq.\ \eqref{annd}]. In the bottom right panel we plot the trajectory of the time evolution in the $(\rho,\phi)$ plane. Note that the points on this trajectory correspond to the integer times $t=0,1, 2,\dots$. We also plot the linear stability threshold for this system. When the trajectory dips below this line, the system temporarily ``melts''.
	}
	\label{68_1}
\end{figure}

\begin{figure}
	\begin {center}
	\includegraphics[width=.5\textwidth]{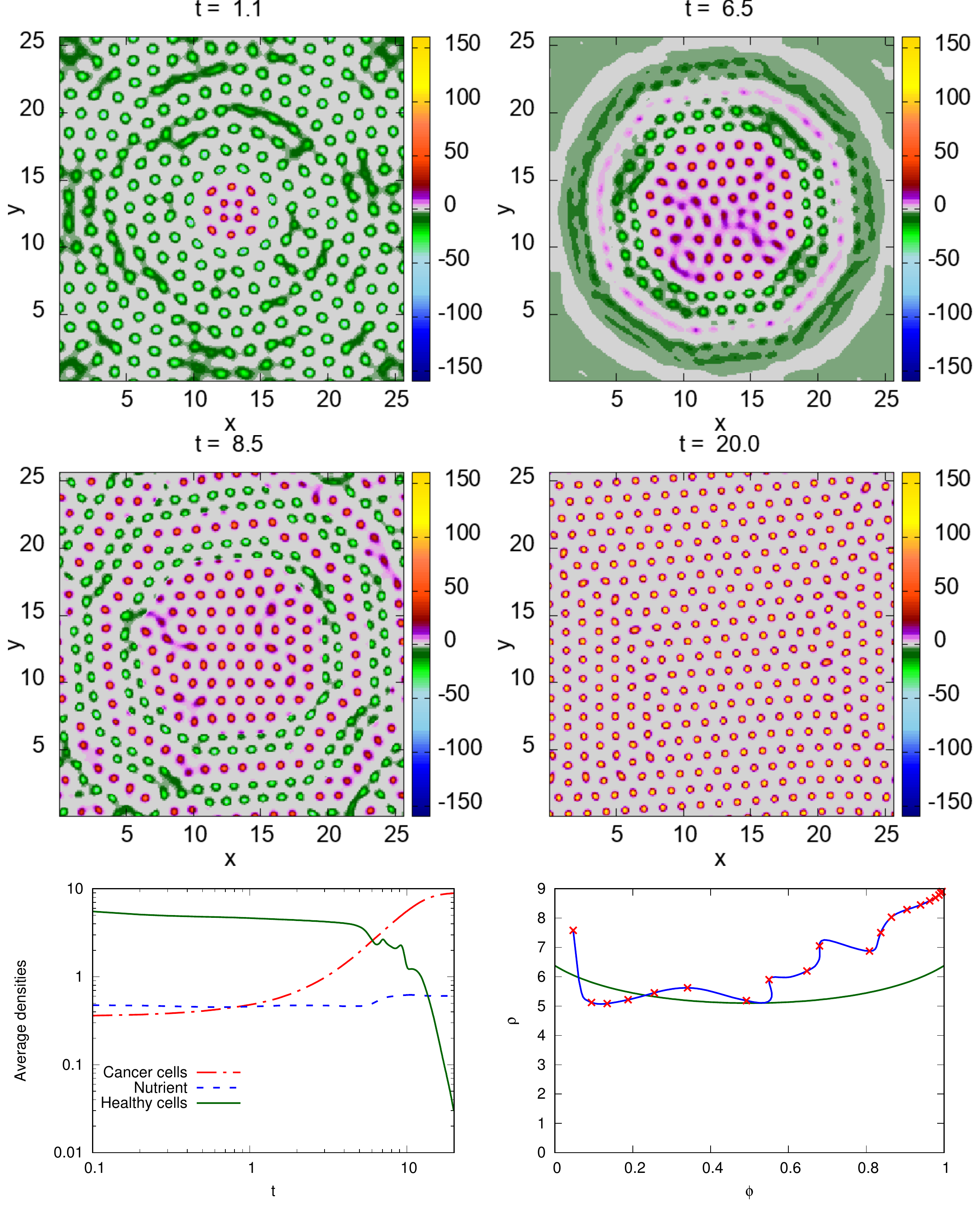}
	\end{center}
	\caption{
	Snapshots of $(\rho_1-\rho_2)$ at the times $t=1.1$, 6.5, 8.5 and 20. All the parameters here are the same as those in Fig.\ \ref{68_1}, except here the cross interaction pair potential radius is $R_{12}=1$, which is slightly larger.
	}
	\label{67}
\end{figure}

\begin{figure}
	\begin {center}	
	\includegraphics[width=.5\textwidth]{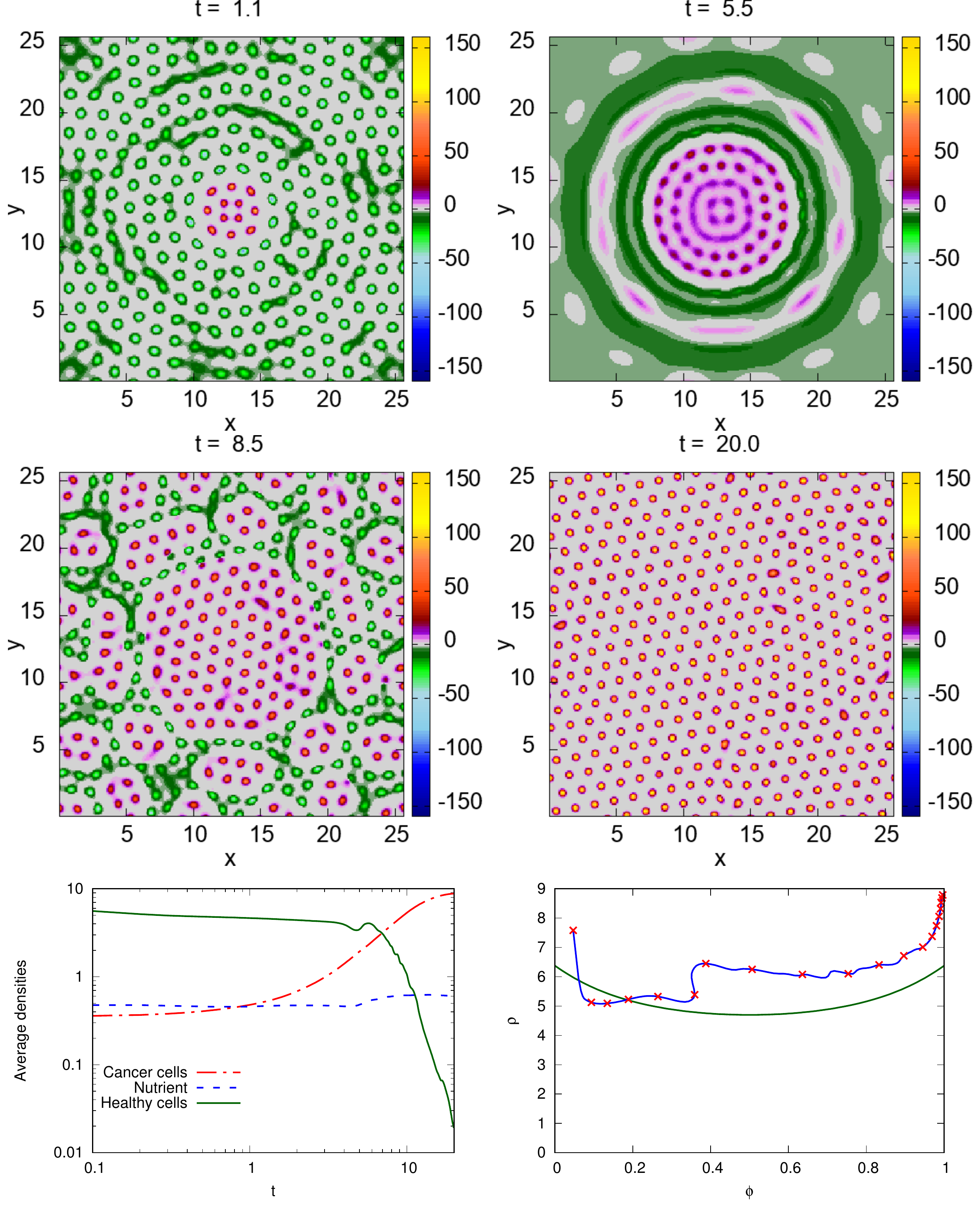}
	\end{center}
	\caption{
	Snapshots of $(\rho_1-\rho_2)$ at the times $t=1.1$, 5.5, 8.5 and 20. All the parameters here are the same as those in Figs.\ \ref{68_1} and \ref{67}, except here the cross interaction pair potential radius is even larger, $R_{12}=1.1$.
	}
	\label{69}
\end{figure}

\correction{Figures \ref{68_1}-\ref{E_ij_L}} \cutout{ In Fig.\ \ref{68_1} we} display results for the evolution over time starting from the initial condition
\begin {equation}\label{IC3}
\rho_{1}(\textbf{r},0)=
 \begin{cases} 
      6+\gamma (\textbf{r}) &\quad (x-12.8)^2+(y-12.8)^2 \le 6^2 \\
      0 &\quad (x-12.8)^2+(y-12.8)^2 > 6^2,
   \end{cases} 
 \end {equation}
 \begin {equation}\label{IC4}
 \rho_{2}(\textbf{r},0)=
 \begin{cases} 
      0 &\quad (x-12.8)^2+(y-12.8)^2 \le 6^2 \\
      6+\gamma (\textbf{r}) &\quad (x-12.8)^2+(y-12.8)^2 > 6^2 
   \end{cases}
\end{equation}
and $n(\textbf{r},0)=0.5$, where $\gamma (\textbf{r})$ is a random variable drawn from a uniform distribution on the interval $(0,1)$. This initial condition corresponds to a small circular cancer of radius 6 in the middle of the healthy cells. \correction{Figs \ref{68_1}-\ref{69} shows simulations with $R_{12}=0.9, 1, 1.1$, respectively, with all other parameters fixed as in Fig \ref{76}, noting that $R_{11}=R_{22}=1$. In the case of $R_{12}=0.9$, the two cell types can tolerate being closer to each other thereby promoting mixing behaviour; this despite the repulsive strength across types, $\beta \varepsilon_{12}=1.5$, being stronger than that between them $\beta \varepsilon_{11}=\beta \varepsilon_{22}=1$. For $R_{12}=1.1$ we expect more demixing type behaviour.} \cutout{ We set the model parameters to be the same as previously.}

We see in Fig.\ \ref{68_1} that although within the domains where the different cell species are initiated -- see Eqs.\ \eqref {IC3} and \eqref{IC4} -- the densities are uniform, i.e.\ liquid--like, rather than a ``crystalline'' state with density peaks, the peaks corresponding to the locations of the cells rapidly form and are already present by the time $t=0.1$. However, this sudden initial growth leads to a drop in the nutrient level, as can be seen at $t \approx$ 5 in Fig.\ \ref{68_1}. The drop in nutrient level then leads to a drop in the overall number of healthy cells, which leads to the ``crystal" melting temporarily, which corresponds to the cells being distributed in disordered liquid-like configurations; biologically, this melting phenomena can be viewed as a temporary state of flux, whereby cells are moving around relatively rapidly and the densities shown are the average density distribution of the cell centres. The nutrient level then recovers and the system ``refreezes'' and over time the cancer cells penetrate the healthy tissue and eventually the healthy cells all die out. This melting phenomenon can be viewed as a state of flux in the system with cells moving around relatively rapidly, thereby the densities shown are more of an average location of the cell centres. The temporary ``melting'' can be understood if one plots the trajectory of the system in the total density versus concentration $(\rho,\phi)$ plane, in addition to plotting the threshold for the system to be linearly unstable, given by Eq.\ \eqref{dk}. This is displayed in the bottom right panel of Fig.\ \ref{68_1}. Recall that above the stability line the system is linearly unstable and forms peaks. We see that when the trajectory dips below this line is when the system temporarily ``melts''.

In the Fig.\ \ref{67} we plot results for the case when all the model parameters are the same as those in the previous case (that displayed in Fig.\ \ref{68_1}), except now the radius in the cross interaction pair potential $R_{12}=1$, which is slightly larger (for the results in Fig.\ \ref{68_1} we have $R_{12}=0.9$). In Fig.\ \ref{67} we plot $(\rho_1-\rho_2)$ at the times $ t$=0.1, 5.5, 9 and 20. As before, we see that the total density of the cancer cells increase with the time and the healthy cells retreat from the centre and finally all the healthy cells die by the time $t=20$. The consequence of the increased value of $R_{12}$ is that there is now a tendency for the cancer cells to penetrate into layers beyond the initial interfacial layer of healthy cells, and so form alternating layers of healthy and cancerous cells -- see e.g.\ the plot for the time $t=7.5$. The averages densities over time are shown in the bottom left panel of the Fig.\ \ref{67} and in the bottom right is the trajectory in the $(\rho,\phi)$ plane and also the corresponding linear stability threshold line.

In Fig.\ \ref{69} we present results for an even larger value of the cross interaction radius, $R_{12}=1.1$. Comparing with Figs.\ \ref{68_1} and \ref{67}, we see that the effect of this increase is to further increase the tendency of the cancer cells to penetrate into the healthy tissue (metastasis) and in this case forming roughly circular clumps of cancer cells ahead of the main tumour, rather than layers.

\correction{The dynamics shown in each of Figs. \ref{68_1}-\ref{69} reflects metastasis. Smaller cross species interaction range, $R_{12}$, lead to a disordered infiltration of healthy tissue by individual tumour cells, which is more ordered for $R_{12}=1$. For the larger $R_{12}$, tumour cells appears to infiltrate healthy tissue as small clusters. In each case, much of the initial mixing of cell types occurs during the transient melting phase, the timescale for which decreases on increasing $R_{12}$ (as can be seen from linear stability threshold diagrams for each of the plots); we note, however, the central core structure of tumour cells is maintained during the melting phase. The different manner of infiltration is an interesting consequence of the modelling assumptions, but it would be experimentally challenging to discern which of these patterns, if any, are relevant biologically. }

\subsubsection{The effect of  varying $\beta \varepsilon_{12}$}

\begin{figure*}
	\begin {center}
	\includegraphics[width=.9\textwidth]{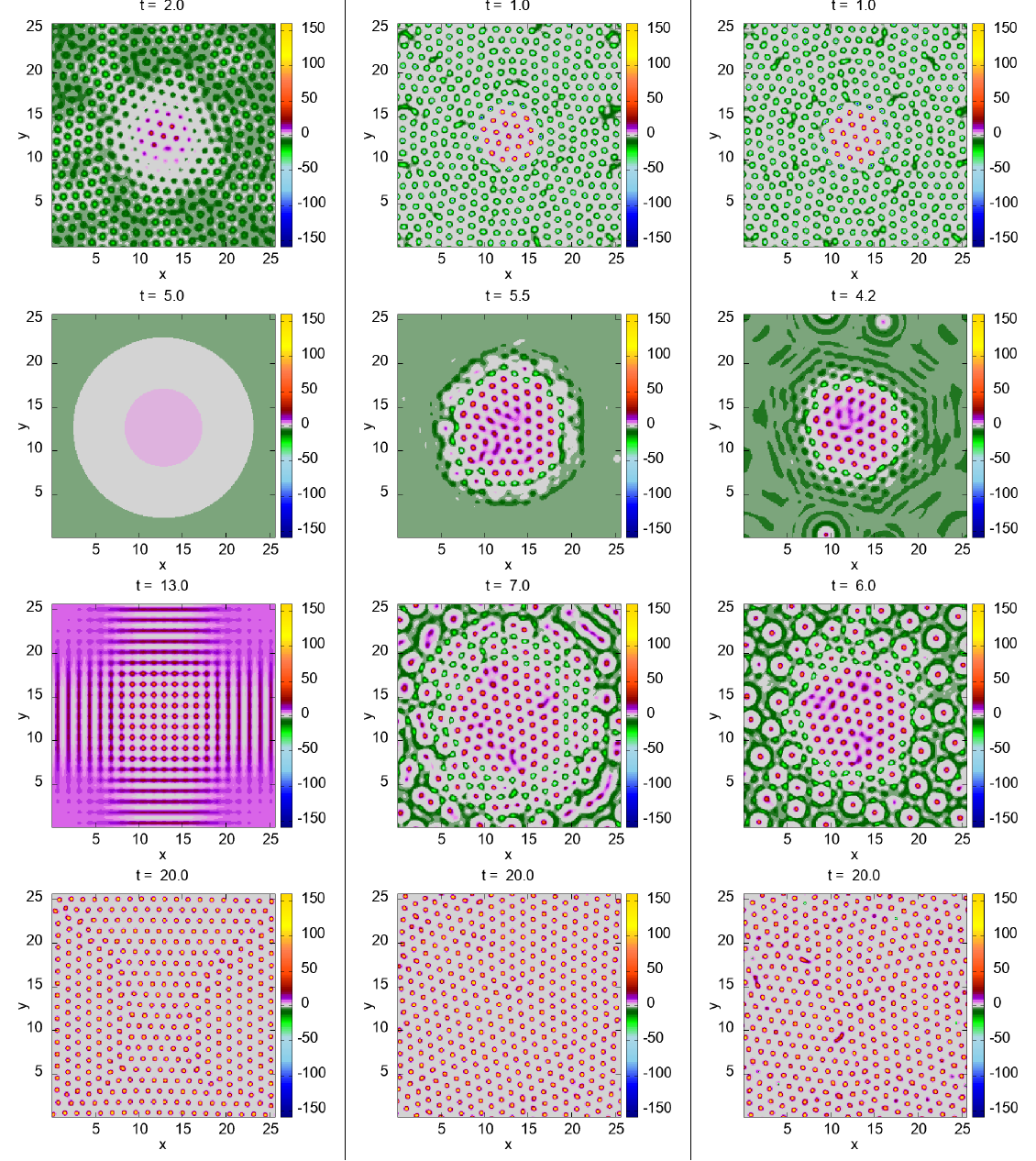}
	\end{center}
	\caption{
	Snapshots of $(\rho_1-\rho_2)$, for various $\beta \varepsilon_{12}=1$ (left), $\beta \varepsilon_{12}=1.75$ (middle) and $\beta \varepsilon_{12}=2$ (right) and various different times, with time increasing from top to bottom, as indicated above. The other pair potential parameters are $\beta \varepsilon_{11}=\beta \varepsilon_{22}=1$, $R_{11}=R_{22}=1$ and $R_{12}=0.9$. The other model parameters are $\tilde \lambda_{n1}=\tilde \lambda_{n2}=1$, $c_1=c_2=0.5$, $\alpha=2$, and $\tilde S_n=9$ with $f(\textbf{r})=1$. The area of the domain is $25.6 \times 25.6$ and $\Delta x=\Delta y=0.1$.
	}
	\label{eij_LT}
\end{figure*}
\begin{figure}
	\begin {center}
	\includegraphics[width=.5\textwidth]{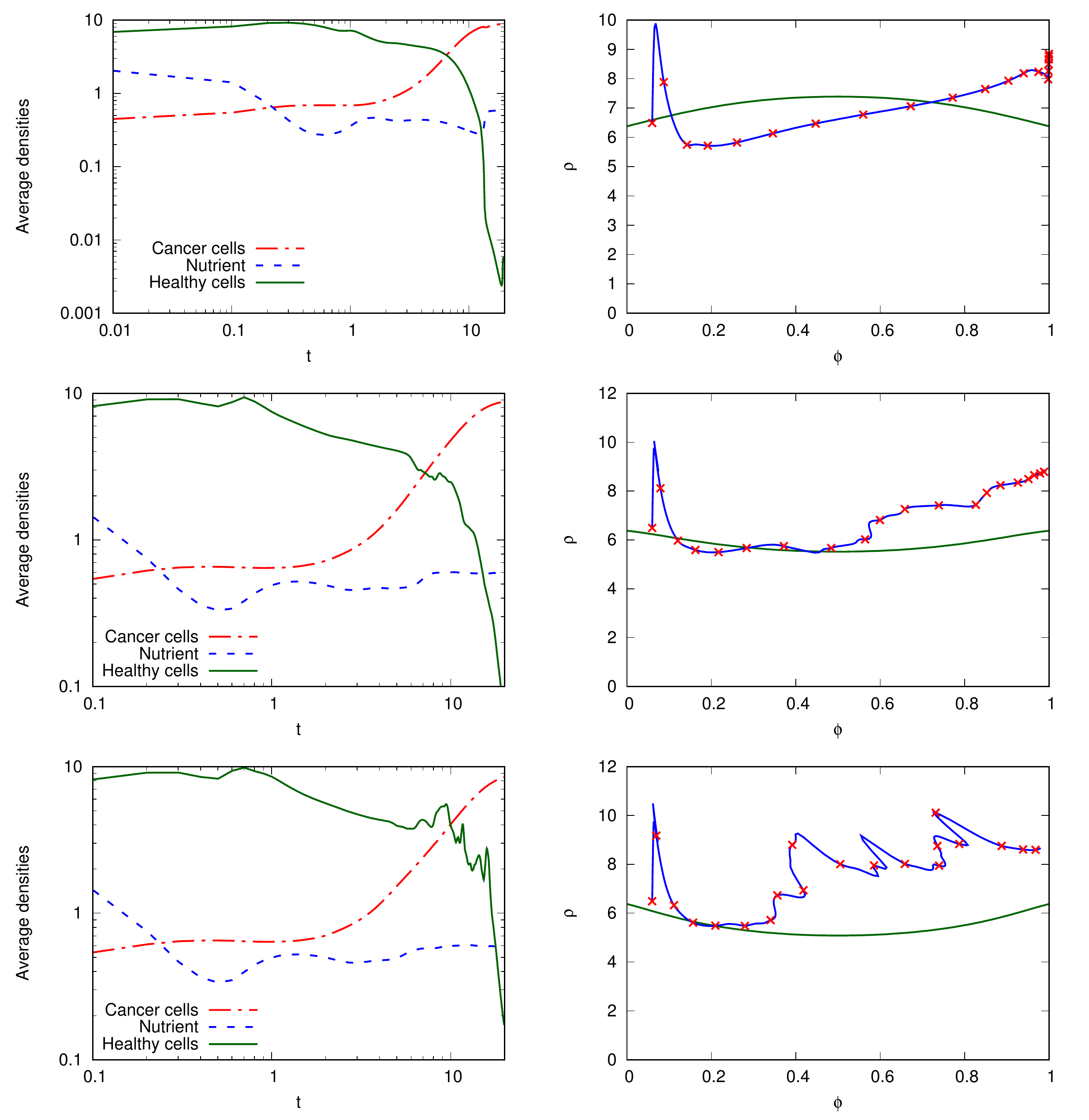}
	\end{center}
	\caption{
	On the left are plots of the average cell densities [see Eq.\ \eqref{acd}] and the average nutrient density [see Eq.\ \eqref{annd}] and on the right plots of the trajectory in the $(\rho,\phi)$ plane with the corresponding linear stability threshold line, corresponding to the results in Fig.\ \ref{eij_LT}. These are for varying $\beta \varepsilon_{12}=1$ (top), $\beta \varepsilon_{12}=1.75$ (middle) and $\beta \varepsilon_{12}=2$ (bottom).
	}
	\label{E_ij_L}
\end{figure}

Guided by the results in Fig.\ \ref{stability2}, we now investigate the effect on the cancer development of varying the cross-species repulsion strength, $\beta \varepsilon_{12}$. In Fig.\ \ref{eij_LT} we display results for three different values, $\beta \varepsilon_{12}=1$, 1.75 and 2. We see that the speed of the cancer cells to penetrate the healthy tissue increases as we increase the value $\beta \varepsilon_{12}$. For the results in the left hand column, which are for $\beta \varepsilon_{12}=1$, there is no penetration of cancer cells into the healthy tissue. For $\beta \varepsilon_{12}=1.75$ (middle column) the penetration starts at $t\approx5.5$ whereas it begins at $t\approx4.5$ for $\beta \varepsilon_{12}=2$ (right hand column).

In Fig.\ \ref{E_ij_L} we plot the average densities of the cells and the nutrient as a function of time and also the trajectory of the system in the $(\rho,\phi)$ plane, corresponding to the results displayed in Fig.\ \ref{eij_LT}. This allows to see that the increased degree of ``melting'' at times $t\sim$ O(1) for smaller $\beta \varepsilon_{12}$ (particularly in the case with $\beta \varepsilon_{12}=1$), is due to the fact that the linear stability threshold line is at higher total densities and is closer to the initial state. This means that the system spends a greater amount of time below the linear stability threshold line as it evolves along its trajectory in the $(\rho,\phi)$ plane. We also see from the plots of the average cell densities over time that the fluctuations over time in the density of the healthy cells increases with increasing $\beta \varepsilon_{12}$. In the $(\rho,\phi)$ plane, these fluctuations manifest as a meandering trajectory with zig-zag-like portions. 

Repeating the simulations corresponding to the results in Figs.\ \ref{eij_LT} and \ref{E_ij_L}, but using $R_{12}=1.1$, such that $R_{12} > \frac{1}{2} (R_{11}+R_{22})$, and also $R_{12}=1$, such that $R_{12}= \frac{1}{2} (R_{11}+R_{22})$, (results not displayed), we find that the results are qualitatively similar, but the melting phenomenon for $\beta \varepsilon_{12}=1$ is prolonged for the smaller value of $R_{12}$ and shortened for the larger value of $R_{12}$. Also, the time at which the cancer cells penetrating into the healthy tissue first appear is earlier for larger $R_{12}$.

\section{Conclusions}

\correction{In this paper we have incorporated DDFT to describe microscopic cell-cell interactions within a simple model of nutrient driven tissue growth.} \cutout{ In this paper, we have developed and studied a continuum model of nutrient driven tissue growth that incorporates cell-cell interactions on a microscopic level using DDFT.} The theory was applied for a single cells type (section \ref{sec1}) and for  two cell types (section \ref{sec61}), the latter representing, for example, the interaction between healthy and tumour cells; this approach can easily be generalised to describe more cells. The resulting models consist of coupled integro-partial differential equations with nonlinear source terms describing nutrient driven growth. This level of description is common in discrete models, but their analysis is limited mainly to numerical simulation; one of the main advantages of the DDFT approach is that the model is amenable to mathematical analysis, providing greater insights into the nature of the numerical results. For instance, the linear stability analysis of Secs.\ \ref{LSA} and \ref{sec:LSA_2} identify parameter regimes for which stable peaks arise, representing the locations of cell centres, as demonstrated in the simulations in Secs.\ \ref{nm} and \ref{sec:num_res_2}. Whilst some parameters can be estimated readily from the experimental literature, this analysis also goes some way to estimate the DDFT associated parameters that are difficult to determine from direct measurements (e.g.\ the effective cell-cell cross interaction radius $R_{12}$). A further outcome of our linear stability analysis in competition case, is the observation that as the cell radii ratio $R_{22}/R_{11}$ is increased, the two wavenumbers at which the system can become linearly unstable move apart leading to the linear stability threshold to develop a cusp. If the radii ratio is sufficiently large (a regime not explored in detail here) then the system can be linearly unstable at two quite different wavenumbers and the interaction between these can produce a wide range of different structures \cite{archer2014solidification,ARK13,archer2016generation} which are interesting from the pattern-formation perspective, and may also have some biological relevance.

\correction{There is still much required in the development of the basic theory before it can be applied directly to experimental results. However, the} \cutout{ The} numerical results reflect qualitatively the expected results based on observation, despite the use of simple growth kinetics and \cutout{ simple}  interaction potentials. For example, the mean densities (a proxy for total number of cells) in Figs.\ \ref{rhogsn} and \ref{rho100} qualitatively resemble Gompertzian or logistic type growth curves often reported in tumour growth models \cite{maruvsic1994tumor}. A further noteworthy aspect of the model is the splitting events shown in Fig.\ \ref{result23}, reflecting mitosis. We note also that for a uniform nutrient distribution, such events are not observed at very large times as the arrangement of the cells settles to fixed configuration; such results are reflective of the cellular rest states observed in mature liver and muscle tissues.

In the simulations of Sec.\ \ref{sec61}, the parameter values for the kinetics guarantee that the tumour cells will overrun the healthy cells. However, it is interesting that the manner by which this is done depends on the value of the interaction parameters $R_{ij}$ and $\epsilon_{ij}$ and in particular the cross-interaction radius $R_{12}$ and energy $\epsilon_{12}$. Although the critical values for $R_{12}$ suggested here are not strictly defined, it was found that (i) if $R_{12}<\frac{1}{2}(R_{11}+R_{22})$, i.e.\ the cross-species interaction range is less then mean of the two same-species interaction ranges, then tumour cells tended to penetrate the healthy regions, whilst (ii) if $R_{12}>\frac{1}{2}(R_{11}+R_{22})$ the tumour cells tend to displace the the healthy cells at the tumour edge, in accordance with the insight gained from studies of mixtures of soft particles \cite{archer2001binary,ALE02,ALE04,GAL06,OvLi09}. Situation (i) is reminiscent of metastasis, whilst (ii) reflects a benign tumour state. Of course, some caution should be applied to such interpretations on the basis of the current analysis, but it is noteworthy that the DDFT approach does identify a potential behavioural property of the cells that can govern benign and virulent tumours. The present work also shows that the overall collective behaviour is sensitive to the details of the pair interactions between cells.

The complex dynamics that the system can exhibit is rather striking. For instance, the drop in the nutrient level observed e.g.\ in Figs.\ \ref{68_1}--\ref{69} that then leads to a drop in the overall number of healthy cells, which results in the ``crystal" melting temporarily, which corresponds to the cells being distributed in disordered liquid-like configurations. The nutrient level then recovers and the system ``refreezes'' and subsequently over time the cancer cells penetrate the healthy tissue and eventually the healthy cells all die out.

The current work is the first to analyse a model using DDFT to describe the growth of tissues and tumours. There is considerable scope to extend the model in order to create a more realistic description of tissue growth. For example, a simple model of EPS was proposed in Ref.\ \cite{CHKLC2012}, whereby EPS gradients generates a haptotactic response of cells, providing a further mechanism for cell movement and arrangement. Another aspect where the present model could be extended relates to the description of the cell-cell interactions. In the models here, these are treated via soft purely repulsive potentials. It would be interesting to compare results with those from alternative soft potential models such as that proposed in Ref.\ \cite{drasdo2007role}. However, in reality there is also attractions (adhesion) between cells, which points to the possibility of the analogue of the gas-liquid or gas-solid phase transitions in collections of cells. Incorporation of both attraction and repulsion between particles in a DFT is straightforward \cite{evans1979nature,Evans92,hansen2013theory}, but the theory becomes much more elaborate, which is why we avoided such theories for this initial study. Despite the current model being very simplistic in comparison to many models of tumour growth, these initial results demonstrate that DDFT has considerable potential as an effective modelling approach to describe microscale cell-cell interactions that can provide new insights into \correction{the dynamics of} tissue and tumour growth. 

\section*{Acknowledgements}

A.A. acknowledges stimulating conversations with John Lowengrub, which helped initiate this work. Hayder Al-Saedi acknowledges the Iraqi Ministry of Higher Education and Scientific Research for financial support.

\label{con}
\appendix
\section{Estimates for parameters values} \label{pv}
  
Here we discuss in further detail what are suitable values to use for the parameters in our model. For the homogeneous system with uniform density, from Eq.\ \eqref{g} we obtain
\begin{equation} 
  \rho(t)=\rho_0 e^{(\lambda_{m}n^{*}-\lambda_d)t},
  \end{equation}
where $\rho_0$ is the initial density. For a given nutrient concentration $n^{*}$ and assuming a 12 hours doubling time \cite{werahera2011proliferative} then from this we can deduce
\begin{equation}\label{zzz}
(\lambda_{m}n^{*} - \lambda_d) = \frac{\ln2}{12} hrs^{-1}.
\end{equation}
 According to  \cite {sherwood1992standard}, a typical value for the concentration of oxygen in fresh water $[O_2]=n^{*}=6.383$ $mg/L$, so we estimate that the critical level $n_d$ for $[O_2]$ is approximately $\frac{n^{*}}{20}=\frac{6.383}{20}=0.32$ $mg/L$ (equivalent to about $1\%$ of atmospheric levels). Hence, $\lambda_m n_d-\lambda_d=0$ leads to 
   \begin{equation}\label{oxx} 
   \lambda_m=\frac{\lambda_d}{0.32  mg/L},
   \end{equation}
and on substitution into Eq.(\ref{zzz}) gives
\begin{equation}\label{}\nonumber
  \lambda_d=0.00005 min^{-1},
  \end{equation}
hence
  \begin{equation}\label{} 
   \lambda_m=0.00015   L min^{-1} mg^{-1}.
   \end{equation}
  The length scale $R$ is the mean radius of the cells, so from Table \ref{table:nonlin2} we have $R \approx10 \mu m$$=0.001 cm$ and in 2 dimensions the typical diffusion distance in time $t$, is estimated from the 2-dimensional average distance diffused squared over time formulae, $\langle r^2 \rangle=4D_c t$. Assuming the time taken to travel a distance of order the diameter of the cell $R$ is about 12 hours, then 
  \begin{equation}\label{} \nonumber
   (2 R)^{2}=4 D_c\times 12 hrs. \Rightarrow  R^{2}=12 D_c 
   \end{equation}
 hence,
   \begin{equation}\label{} \nonumber
  D_c=\frac{R^{2}}{12 hrs.}=\frac{0.001^2}{12 \times 60 min.}=1.3\times 10^{-9} cm^2/min
  \end{equation}
The dimensionless population growth constant is $c_1=\frac {R^2 \lambda_d}{D_c}$, so we get $c_1=0.038$. From the definition of $\tilde D_c =\frac {D_n}{D_c}$, and $D_n=2 \times 10^{-5} cm^2/sec$ ($D_n=1.2 \times 10^{-3} cm^2/min$) \cite{ward1999mathematical,ward1999math}, this leads to
\begin{equation}\label{diffusion}
  \tilde D =\frac {12 \times10^{-4}}{13 \times 10^{-10}}\approx1 \times10^{6}.
  \end{equation}
The nutrient source term $\tilde S_n=\frac{R^2 S_n \lambda_m}{\lambda_d D_c}$ is estimated to be O$(10^{6})$ so that in Eq. \eqref{second} $\bar n$ is in balance with the diffusion term. Hence $ \frac {3}{13}\times10^{4} S_n\approx(10^{6})$  $\Rightarrow S_n=433 $. From Eq. \eqref{second} we also see that the term involving $ \tilde \lambda_n$ also must 
 balance with diffusion, hence from Eq. \eqref{lamda} we see $\lambda_n$  must be O$(10^{-4})$ to ensure that $\tilde \lambda_n$ is of O$(10^6)$. 
Recall that the number density is the number of cells per unit area $\rho=\frac{N}{A}$. Since $R \approx10 \mu m$ $=0.001 \, cm$, this implies that the area covered by one circular cell $=\pi R^2 \approx 3 \times 10 ^{-6} cm^2$. This then implies that a typical cell density is $\rho \approx \frac{1}{3} \times 10^{6}  cm^{-2}$ i.e. $3 \times 10^{5} cm^{-2}$.

We summarise the values of dimensional parameters in Table \ref{table:nonlin2} and dimensionless parameter values in Table \ref{table:nonlin}.


\begin{thebibliography}{75}%
\makeatletter
\providecommand \@ifxundefined [1]{%
 \@ifx{#1\undefined}
}%
\providecommand \@ifnum [1]{%
 \ifnum #1\expandafter \@firstoftwo
 \else \expandafter \@secondoftwo
 \fi
}%
\providecommand \@ifx [1]{%
 \ifx #1\expandafter \@firstoftwo
 \else \expandafter \@secondoftwo
 \fi
}%
\providecommand \natexlab [1]{#1}%
\providecommand \enquote  [1]{``#1''}%
\providecommand \bibnamefont  [1]{#1}%
\providecommand \bibfnamefont [1]{#1}%
\providecommand \citenamefont [1]{#1}%
\providecommand \href@noop [0]{\@secondoftwo}%
\providecommand \href [0]{\begingroup \@sanitize@url \@href}%
\providecommand \@href[1]{\@@startlink{#1}\@@href}%
\providecommand \@@href[1]{\endgroup#1\@@endlink}%
\providecommand \@sanitize@url [0]{\catcode `\\12\catcode `\$12\catcode
  `\&12\catcode `\#12\catcode `\^12\catcode `\_12\catcode `\%12\relax}%
\providecommand \@@startlink[1]{}%
\providecommand \@@endlink[0]{}%
\providecommand \url  [0]{\begingroup\@sanitize@url \@url }%
\providecommand \@url [1]{\endgroup\@href {#1}{\urlprefix }}%
\providecommand \urlprefix  [0]{URL }%
\providecommand \Eprint [0]{\href }%
\providecommand \doibase [0]{http://dx.doi.org/}%
\providecommand \selectlanguage [0]{\@gobble}%
\providecommand \bibinfo  [0]{\@secondoftwo}%
\providecommand \bibfield  [0]{\@secondoftwo}%
\providecommand \translation [1]{[#1]}%
\providecommand \BibitemOpen [0]{}%
\providecommand \bibitemStop [0]{}%
\providecommand \bibitemNoStop [0]{.\EOS\space}%
\providecommand \EOS [0]{\spacefactor3000\relax}%
\providecommand \BibitemShut  [1]{\csname bibitem#1\endcsname}%
\let\auto@bib@innerbib\@empty
\bibitem [{\citenamefont {Byrne}(2000)}]{byrne2000using}%
  \BibitemOpen
  \bibfield  {author} {\bibinfo {author} {\bibfnamefont {H.~M.}\ \bibnamefont
  {Byrne}},\ }in\ \href@noop {} {\emph {\bibinfo {booktitle} {Proceedings of
  the 9th General Meetings of European Women in Mathematics}}}\ (\bibinfo
  {year} {2000})\ pp.\ \bibinfo {pages} {81--107}\BibitemShut {NoStop}%
\bibitem [{\citenamefont {Siegel}\ \emph {et~al.}(2012)\citenamefont {Siegel},
  \citenamefont {DeSantis}, \citenamefont {Virgo}, \citenamefont {Stein},
  \citenamefont {Mariotto}, \citenamefont {Smith}, \citenamefont {Cooper},
  \citenamefont {Gansler}, \citenamefont {Lerro}, \citenamefont {Fedewa} \emph
  {et~al.}}]{siegel2012cancer}%
  \BibitemOpen
  \bibfield  {author} {\bibinfo {author} {\bibfnamefont {R.}~\bibnamefont
  {Siegel}}, \bibinfo {author} {\bibfnamefont {C.}~\bibnamefont {DeSantis}},
  \bibinfo {author} {\bibfnamefont {K.}~\bibnamefont {Virgo}}, \bibinfo
  {author} {\bibfnamefont {K.}~\bibnamefont {Stein}}, \bibinfo {author}
  {\bibfnamefont {A.}~\bibnamefont {Mariotto}}, \bibinfo {author}
  {\bibfnamefont {T.}~\bibnamefont {Smith}}, \bibinfo {author} {\bibfnamefont
  {D.}~\bibnamefont {Cooper}}, \bibinfo {author} {\bibfnamefont
  {T.}~\bibnamefont {Gansler}}, \bibinfo {author} {\bibfnamefont
  {C.}~\bibnamefont {Lerro}}, \bibinfo {author} {\bibfnamefont
  {S.}~\bibnamefont {Fedewa}},  \emph {et~al.},\ }\href@noop {} {\bibfield
  {journal} {\bibinfo  {journal} {CA: A Cancer Journal for Clinicians}\
  }\textbf {\bibinfo {volume} {62}},\ \bibinfo {pages} {220} (\bibinfo {year}
  {2012})}\BibitemShut {NoStop}%
\bibitem [{\citenamefont {Weinberg}(2013)}]{weinberg2013biology}%
  \BibitemOpen
  \bibfield  {author} {\bibinfo {author} {\bibfnamefont {R.}~\bibnamefont
  {Weinberg}},\ }\href@noop {} {\emph {\bibinfo {title} {The biology of
  cancer}}}\ (\bibinfo  {publisher} {Garland science},\ \bibinfo {year}
  {2013})\BibitemShut {NoStop}%
\bibitem [{\citenamefont {Burton}(1966)}]{burton1966rate}%
  \BibitemOpen
  \bibfield  {author} {\bibinfo {author} {\bibfnamefont {A.~C.}\ \bibnamefont
  {Burton}},\ }\href@noop {} {\bibfield  {journal} {\bibinfo  {journal}
  {Growth}\ }\textbf {\bibinfo {volume} {30}},\ \bibinfo {pages} {157}
  (\bibinfo {year} {1966})}\BibitemShut {NoStop}%
\bibitem [{\citenamefont {Greenspan}(1972)}]{greenspan1972models}%
  \BibitemOpen
  \bibfield  {author} {\bibinfo {author} {\bibfnamefont {H.}~\bibnamefont
  {Greenspan}},\ }\href@noop {} {\bibfield  {journal} {\bibinfo  {journal}
  {Stud. Appl. Math.}\ }\textbf {\bibinfo {volume} {51}},\ \bibinfo {pages}
  {317} (\bibinfo {year} {1972})}\BibitemShut {NoStop}%
\bibitem [{\citenamefont {Sutherland}\ \emph {et~al.}(1971)\citenamefont
  {Sutherland}, \citenamefont {McCredie},\ and\ \citenamefont
  {Inch}}]{sutherland1971growth}%
  \BibitemOpen
  \bibfield  {author} {\bibinfo {author} {\bibfnamefont {R.~M.}\ \bibnamefont
  {Sutherland}}, \bibinfo {author} {\bibfnamefont {J.~A.}\ \bibnamefont
  {McCredie}}, \ and\ \bibinfo {author} {\bibfnamefont {W.~R.}\ \bibnamefont
  {Inch}},\ }\href@noop {} {\bibfield  {journal} {\bibinfo  {journal} {Journal
  of the National Cancer Institute}\ }\textbf {\bibinfo {volume} {46}},\
  \bibinfo {pages} {113} (\bibinfo {year} {1971})}\BibitemShut {NoStop}%
\bibitem [{\citenamefont {Glass}(1973)}]{glass1973instability}%
  \BibitemOpen
  \bibfield  {author} {\bibinfo {author} {\bibfnamefont {L.}~\bibnamefont
  {Glass}},\ }\href@noop {} {\bibfield  {journal} {\bibinfo  {journal} {Journal
  of Dynamic Systems, Measurement, and Control}\ }\textbf {\bibinfo {volume}
  {95}},\ \bibinfo {pages} {324} (\bibinfo {year} {1973})}\BibitemShut
  {NoStop}%
\bibitem [{\citenamefont {Bullough}(1965)}]{bullough1965mitotic}%
  \BibitemOpen
  \bibfield  {author} {\bibinfo {author} {\bibfnamefont {W.~S.}\ \bibnamefont
  {Bullough}},\ }\href@noop {} {\bibfield  {journal} {\bibinfo  {journal}
  {Cancer Research}\ }\textbf {\bibinfo {volume} {25}},\ \bibinfo {pages}
  {1683} (\bibinfo {year} {1965})}\BibitemShut {NoStop}%
\bibitem [{\citenamefont {Ward}\ and\ \citenamefont
  {King}(1997)}]{ward1997mathematical}%
  \BibitemOpen
  \bibfield  {author} {\bibinfo {author} {\bibfnamefont {J.~P.}\ \bibnamefont
  {Ward}}\ and\ \bibinfo {author} {\bibfnamefont {J.}~\bibnamefont {King}},\
  }\href@noop {} {\bibfield  {journal} {\bibinfo  {journal} {Mathematical
  Medicine and Biology}\ }\textbf {\bibinfo {volume} {14}},\ \bibinfo {pages}
  {39} (\bibinfo {year} {1997})}\BibitemShut {NoStop}%
\bibitem [{\citenamefont {Lowengrub}\ \emph {et~al.}(2009)\citenamefont
  {Lowengrub}, \citenamefont {Frieboes}, \citenamefont {Jin}, \citenamefont
  {Chuang}, \citenamefont {Li}, \citenamefont {Macklin}, \citenamefont {Wise},\
  and\ \citenamefont {Cristini}}]{lowengrub2009nonlinear}%
  \BibitemOpen
  \bibfield  {author} {\bibinfo {author} {\bibfnamefont {J.~S.}\ \bibnamefont
  {Lowengrub}}, \bibinfo {author} {\bibfnamefont {H.~B.}\ \bibnamefont
  {Frieboes}}, \bibinfo {author} {\bibfnamefont {F.}~\bibnamefont {Jin}},
  \bibinfo {author} {\bibfnamefont {Y.}~\bibnamefont {Chuang}}, \bibinfo
  {author} {\bibfnamefont {X.}~\bibnamefont {Li}}, \bibinfo {author}
  {\bibfnamefont {P.}~\bibnamefont {Macklin}}, \bibinfo {author} {\bibfnamefont
  {S.~M.}\ \bibnamefont {Wise}}, \ and\ \bibinfo {author} {\bibfnamefont
  {V.}~\bibnamefont {Cristini}},\ }\href@noop {} {\bibfield  {journal}
  {\bibinfo  {journal} {Nonlinearity}\ }\textbf {\bibinfo {volume} {23}},\
  \bibinfo {pages} {R1} (\bibinfo {year} {2009})}\BibitemShut {NoStop}%
\bibitem [{\citenamefont {Adam}(1986)}]{adam1986simplified}%
  \BibitemOpen
  \bibfield  {author} {\bibinfo {author} {\bibfnamefont {J.~A.}\ \bibnamefont
  {Adam}},\ }\href@noop {} {\bibfield  {journal} {\bibinfo  {journal}
  {Mathematical biosciences}\ }\textbf {\bibinfo {volume} {81}},\ \bibinfo
  {pages} {229} (\bibinfo {year} {1986})}\BibitemShut {NoStop}%
\bibitem [{\citenamefont {McElwain}\ \emph {et~al.}(1979)\citenamefont
  {McElwain}, \citenamefont {Callcott},\ and\ \citenamefont
  {Morris}}]{mcelwain1979model}%
  \BibitemOpen
  \bibfield  {author} {\bibinfo {author} {\bibfnamefont {D.~L.~S.}\
  \bibnamefont {McElwain}}, \bibinfo {author} {\bibfnamefont {R.}~\bibnamefont
  {Callcott}}, \ and\ \bibinfo {author} {\bibfnamefont {L.~E.}\ \bibnamefont
  {Morris}},\ }\href@noop {} {\bibfield  {journal} {\bibinfo  {journal} {J.
  Theor. Biology}\ }\textbf {\bibinfo {volume} {78}},\ \bibinfo {pages} {405}
  (\bibinfo {year} {1979})}\BibitemShut {NoStop}%
\bibitem [{\citenamefont {Breward}\ \emph {et~al.}(2003)\citenamefont
  {Breward}, \citenamefont {Byrne},\ and\ \citenamefont
  {Lewis}}]{breward2003multiphase}%
  \BibitemOpen
  \bibfield  {author} {\bibinfo {author} {\bibfnamefont {C.~J.}\ \bibnamefont
  {Breward}}, \bibinfo {author} {\bibfnamefont {H.~M.}\ \bibnamefont {Byrne}},
  \ and\ \bibinfo {author} {\bibfnamefont {C.~E.}\ \bibnamefont {Lewis}},\
  }\href@noop {} {\bibfield  {journal} {\bibinfo  {journal} {Bull. Math. Bio.}\
  }\textbf {\bibinfo {volume} {65}},\ \bibinfo {pages} {609} (\bibinfo {year}
  {2003})}\BibitemShut {NoStop}%
\bibitem [{\citenamefont {Orme}\ and\ \citenamefont
  {Chaplain}(1996)}]{orme1996mathematical}%
  \BibitemOpen
  \bibfield  {author} {\bibinfo {author} {\bibfnamefont {M.~E.}\ \bibnamefont
  {Orme}}\ and\ \bibinfo {author} {\bibfnamefont {M.~A.~J.}\ \bibnamefont
  {Chaplain}},\ }\href@noop {} {\bibfield  {journal} {\bibinfo  {journal}
  {Mathematical and Computer Modelling}\ }\textbf {\bibinfo {volume} {23}},\
  \bibinfo {pages} {43} (\bibinfo {year} {1996})}\BibitemShut {NoStop}%
\bibitem [{\citenamefont {Byrne}\ and\ \citenamefont
  {Chaplain}(1995)}]{byrne1995growth}%
  \BibitemOpen
  \bibfield  {author} {\bibinfo {author} {\bibfnamefont {H.~M.}\ \bibnamefont
  {Byrne}}\ and\ \bibinfo {author} {\bibfnamefont {M.~A.~J.}\ \bibnamefont
  {Chaplain}},\ }\href@noop {} {\bibfield  {journal} {\bibinfo  {journal}
  {Mathematical Biosciences}\ }\textbf {\bibinfo {volume} {130}},\ \bibinfo
  {pages} {151} (\bibinfo {year} {1995})}\BibitemShut {NoStop}%
\bibitem [{\citenamefont {Kerbel}(2000)}]{kerbel2000tumor}%
  \BibitemOpen
  \bibfield  {author} {\bibinfo {author} {\bibfnamefont {R.~S.}\ \bibnamefont
  {Kerbel}},\ }\href@noop {} {\bibfield  {journal} {\bibinfo  {journal}
  {Carcinogenesis}\ }\textbf {\bibinfo {volume} {21}},\ \bibinfo {pages} {505}
  (\bibinfo {year} {2000})}\BibitemShut {NoStop}%
\bibitem [{\citenamefont {Viallard}\ and\ \citenamefont
  {Larriv{\'e}e}(2017)}]{viallard2017tumor}%
  \BibitemOpen
  \bibfield  {author} {\bibinfo {author} {\bibfnamefont {C.}~\bibnamefont
  {Viallard}}\ and\ \bibinfo {author} {\bibfnamefont {B.}~\bibnamefont
  {Larriv{\'e}e}},\ }\href@noop {} {\bibfield  {journal} {\bibinfo  {journal}
  {Angiogenesis}\ ,\ \bibinfo {pages} {1}} (\bibinfo {year}
  {2017})}\BibitemShut {NoStop}%
\bibitem [{\citenamefont {Kim}\ \emph {et~al.}(2008)\citenamefont {Kim},
  \citenamefont {Lee},\ and\ \citenamefont {Levy}}]{kim2008pde}%
  \BibitemOpen
  \bibfield  {author} {\bibinfo {author} {\bibfnamefont {P.~S.}\ \bibnamefont
  {Kim}}, \bibinfo {author} {\bibfnamefont {P.~P.}\ \bibnamefont {Lee}}, \ and\
  \bibinfo {author} {\bibfnamefont {D.}~\bibnamefont {Levy}},\ }\href@noop {}
  {\bibfield  {journal} {\bibinfo  {journal} {Bull. Math. Bio.}\ }\textbf
  {\bibinfo {volume} {70}},\ \bibinfo {pages} {1994} (\bibinfo {year}
  {2008})}\BibitemShut {NoStop}%
\bibitem [{\citenamefont {Miller}\ \emph {et~al.}(2016)\citenamefont {Miller},
  \citenamefont {Siegel}, \citenamefont {Lin}, \citenamefont {Mariotto},
  \citenamefont {Kramer}, \citenamefont {Rowland}, \citenamefont {Stein},
  \citenamefont {Alteri},\ and\ \citenamefont {Jemal}}]{miller2016cancer}%
  \BibitemOpen
  \bibfield  {author} {\bibinfo {author} {\bibfnamefont {K.~D.}\ \bibnamefont
  {Miller}}, \bibinfo {author} {\bibfnamefont {R.~L.}\ \bibnamefont {Siegel}},
  \bibinfo {author} {\bibfnamefont {C.~C.}\ \bibnamefont {Lin}}, \bibinfo
  {author} {\bibfnamefont {A.~B.}\ \bibnamefont {Mariotto}}, \bibinfo {author}
  {\bibfnamefont {J.~L.}\ \bibnamefont {Kramer}}, \bibinfo {author}
  {\bibfnamefont {J.~H.}\ \bibnamefont {Rowland}}, \bibinfo {author}
  {\bibfnamefont {K.~D.}\ \bibnamefont {Stein}}, \bibinfo {author}
  {\bibfnamefont {R.}~\bibnamefont {Alteri}}, \ and\ \bibinfo {author}
  {\bibfnamefont {A.}~\bibnamefont {Jemal}},\ }\href@noop {} {\bibfield
  {journal} {\bibinfo  {journal} {CA: a Cancer Journal for Clinicians}\
  }\textbf {\bibinfo {volume} {66}},\ \bibinfo {pages} {271} (\bibinfo {year}
  {2016})}\BibitemShut {NoStop}%
\bibitem [{\citenamefont {Ward}\ and\ \citenamefont
  {King}(2003)}]{ward2003mathematical}%
  \BibitemOpen
  \bibfield  {author} {\bibinfo {author} {\bibfnamefont {J.~P.}\ \bibnamefont
  {Ward}}\ and\ \bibinfo {author} {\bibfnamefont {J.~R.}\ \bibnamefont
  {King}},\ }\href@noop {} {\bibfield  {journal} {\bibinfo  {journal}
  {Mathematical biosciences}\ }\textbf {\bibinfo {volume} {181}},\ \bibinfo
  {pages} {177} (\bibinfo {year} {2003})}\BibitemShut {NoStop}%
\bibitem [{\citenamefont {Ruoslahti}(1996)}]{ruoslahti1996cancer}%
  \BibitemOpen
  \bibfield  {author} {\bibinfo {author} {\bibfnamefont {E.}~\bibnamefont
  {Ruoslahti}},\ }\href@noop {} {\bibfield  {journal} {\bibinfo  {journal}
  {Scientific American}\ }\textbf {\bibinfo {volume} {275}},\ \bibinfo {pages}
  {72} (\bibinfo {year} {1996})}\BibitemShut {NoStop}%
\bibitem [{\citenamefont {Liotta}\ \emph {et~al.}(1974)\citenamefont {Liotta},
  \citenamefont {Kleinerman},\ and\ \citenamefont
  {Saidel}}]{liotta1974quantitative}%
  \BibitemOpen
  \bibfield  {author} {\bibinfo {author} {\bibfnamefont {L.~A.}\ \bibnamefont
  {Liotta}}, \bibinfo {author} {\bibfnamefont {J.}~\bibnamefont {Kleinerman}},
  \ and\ \bibinfo {author} {\bibfnamefont {G.~M.}\ \bibnamefont {Saidel}},\
  }\href@noop {} {\bibfield  {journal} {\bibinfo  {journal} {Cancer research}\
  }\textbf {\bibinfo {volume} {34}},\ \bibinfo {pages} {997} (\bibinfo {year}
  {1974})}\BibitemShut {NoStop}%
\bibitem [{\citenamefont {Liotta}\ \emph {et~al.}(1976)\citenamefont {Liotta},
  \citenamefont {Kleinerman},\ and\ \citenamefont
  {Saldel}}]{liotta1976significance}%
  \BibitemOpen
  \bibfield  {author} {\bibinfo {author} {\bibfnamefont {L.~A.}\ \bibnamefont
  {Liotta}}, \bibinfo {author} {\bibfnamefont {J.}~\bibnamefont {Kleinerman}},
  \ and\ \bibinfo {author} {\bibfnamefont {G.~M.}\ \bibnamefont {Saldel}},\
  }\href@noop {} {\bibfield  {journal} {\bibinfo  {journal} {Cancer research}\
  }\textbf {\bibinfo {volume} {36}},\ \bibinfo {pages} {889} (\bibinfo {year}
  {1976})}\BibitemShut {NoStop}%
\bibitem [{\citenamefont {Adam}(1987)}]{adam1987mathematical}%
  \BibitemOpen
  \bibfield  {author} {\bibinfo {author} {\bibfnamefont {J.~A.}\ \bibnamefont
  {Adam}},\ }\href@noop {} {\bibfield  {journal} {\bibinfo  {journal}
  {Mathematical Biosciences}\ }\textbf {\bibinfo {volume} {86}},\ \bibinfo
  {pages} {183} (\bibinfo {year} {1987})}\BibitemShut {NoStop}%
\bibitem [{\citenamefont {Chen}\ and\ \citenamefont
  {Ward}(2014)}]{chen2014mathematical}%
  \BibitemOpen
  \bibfield  {author} {\bibinfo {author} {\bibfnamefont {C.-Y.}\ \bibnamefont
  {Chen}}\ and\ \bibinfo {author} {\bibfnamefont {J.~P.}\ \bibnamefont
  {Ward}},\ }\href@noop {} {\bibfield  {journal} {\bibinfo  {journal} {Bulletin
  of mathematical biology}\ }\textbf {\bibinfo {volume} {76}},\ \bibinfo
  {pages} {3088} (\bibinfo {year} {2014})}\BibitemShut {NoStop}%
\bibitem [{\citenamefont {Pierres}\ \emph {et~al.}(2000)\citenamefont
  {Pierres}, \citenamefont {Benoliel},\ and\ \citenamefont
  {Bongrand}}]{pierres2000cell}%
  \BibitemOpen
  \bibfield  {author} {\bibinfo {author} {\bibfnamefont {A.}~\bibnamefont
  {Pierres}}, \bibinfo {author} {\bibfnamefont {A.~M.}\ \bibnamefont
  {Benoliel}}, \ and\ \bibinfo {author} {\bibfnamefont {P.}~\bibnamefont
  {Bongrand}},\ }\href@noop {} {\emph {\bibinfo {title} {{\it Cell-cell
  interactions}, in Physical Chemistry of Biological Interfaces}}}\ (\bibinfo
  {publisher} {Marcel Dekker, New York},\ \bibinfo {year} {2000})\ pp.\
  \bibinfo {pages} {459--522}\BibitemShut {NoStop}%
\bibitem [{\citenamefont {Piotrowska}\ and\ \citenamefont
  {Angus}(2009)}]{piotrowska2009quantitative}%
  \BibitemOpen
  \bibfield  {author} {\bibinfo {author} {\bibfnamefont {M.~J.}\ \bibnamefont
  {Piotrowska}}\ and\ \bibinfo {author} {\bibfnamefont {S.~D.}\ \bibnamefont
  {Angus}},\ }\href@noop {} {\bibfield  {journal} {\bibinfo  {journal} {J.
  Theor. Biology}\ }\textbf {\bibinfo {volume} {258}},\ \bibinfo {pages} {165}
  (\bibinfo {year} {2009})}\BibitemShut {NoStop}%
\bibitem [{\citenamefont {Anderson}(2005)}]{anderson2005hybrid}%
  \BibitemOpen
  \bibfield  {author} {\bibinfo {author} {\bibfnamefont {A.~R.~A.}\
  \bibnamefont {Anderson}},\ }\href@noop {} {\bibfield  {journal} {\bibinfo
  {journal} {Mathematical medicine and biology: a journal of the IMA}\ }\textbf
  {\bibinfo {volume} {22}},\ \bibinfo {pages} {163} (\bibinfo {year}
  {2005})}\BibitemShut {NoStop}%
\bibitem [{\citenamefont {Engelberg}\ \emph {et~al.}(2008)\citenamefont
  {Engelberg}, \citenamefont {Ropella},\ and\ \citenamefont
  {Hunt}}]{engelberg2008essential}%
  \BibitemOpen
  \bibfield  {author} {\bibinfo {author} {\bibfnamefont {J.~A.}\ \bibnamefont
  {Engelberg}}, \bibinfo {author} {\bibfnamefont {G.~E.~P.}\ \bibnamefont
  {Ropella}}, \ and\ \bibinfo {author} {\bibfnamefont {C.~A.}\ \bibnamefont
  {Hunt}},\ }\href@noop {} {\bibfield  {journal} {\bibinfo  {journal} {BMC
  Systems Biology}\ }\textbf {\bibinfo {volume} {2}},\ \bibinfo {pages} {110}
  (\bibinfo {year} {2008})}\BibitemShut {NoStop}%
\bibitem [{\citenamefont {Drasdo}\ and\ \citenamefont
  {H{\"o}hme}(2005)}]{drasdo2005single}%
  \BibitemOpen
  \bibfield  {author} {\bibinfo {author} {\bibfnamefont {D.}~\bibnamefont
  {Drasdo}}\ and\ \bibinfo {author} {\bibfnamefont {S.}~\bibnamefont
  {H{\"o}hme}},\ }\href@noop {} {\bibfield  {journal} {\bibinfo  {journal}
  {Physical Biology}\ }\textbf {\bibinfo {volume} {2}},\ \bibinfo {pages} {133}
  (\bibinfo {year} {2005})}\BibitemShut {NoStop}%
\bibitem [{\citenamefont {Galle}\ \emph {et~al.}(2009)\citenamefont {Galle},
  \citenamefont {Hoffmann},\ and\ \citenamefont {Aust}}]{galle2009single}%
  \BibitemOpen
  \bibfield  {author} {\bibinfo {author} {\bibfnamefont {J.}~\bibnamefont
  {Galle}}, \bibinfo {author} {\bibfnamefont {M.}~\bibnamefont {Hoffmann}}, \
  and\ \bibinfo {author} {\bibfnamefont {G.}~\bibnamefont {Aust}},\ }\href@noop
  {} {\bibfield  {journal} {\bibinfo  {journal} {J. Math. Bio.}\ }\textbf
  {\bibinfo {volume} {58}},\ \bibinfo {pages} {261} (\bibinfo {year}
  {2009})}\BibitemShut {NoStop}%
\bibitem [{\citenamefont {Jeon}\ \emph {et~al.}(2010)\citenamefont {Jeon},
  \citenamefont {Quaranta},\ and\ \citenamefont {Cummings}}]{jeon2010off}%
  \BibitemOpen
  \bibfield  {author} {\bibinfo {author} {\bibfnamefont {J.}~\bibnamefont
  {Jeon}}, \bibinfo {author} {\bibfnamefont {V.}~\bibnamefont {Quaranta}}, \
  and\ \bibinfo {author} {\bibfnamefont {P.~T.}\ \bibnamefont {Cummings}},\
  }\href@noop {} {\bibfield  {journal} {\bibinfo  {journal} {Biophys. J.}\
  }\textbf {\bibinfo {volume} {98}},\ \bibinfo {pages} {37} (\bibinfo {year}
  {2010})}\BibitemShut {NoStop}%
\bibitem [{\citenamefont {Turner}\ \emph {et~al.}(2004)\citenamefont {Turner},
  \citenamefont {Sherratt},\ and\ \citenamefont
  {Cameron}}]{turner2004tamoxifen}%
  \BibitemOpen
  \bibfield  {author} {\bibinfo {author} {\bibfnamefont {S.}~\bibnamefont
  {Turner}}, \bibinfo {author} {\bibfnamefont {J.~A.}\ \bibnamefont
  {Sherratt}}, \ and\ \bibinfo {author} {\bibfnamefont {D.}~\bibnamefont
  {Cameron}},\ }\href@noop {} {\bibfield  {journal} {\bibinfo  {journal} {J.
  Theor. Bio.}\ }\textbf {\bibinfo {volume} {229}},\ \bibinfo {pages} {101}
  (\bibinfo {year} {2004})}\BibitemShut {NoStop}%
\bibitem [{\citenamefont {Mar{\'e}e}\ \emph {et~al.}(2007)\citenamefont
  {Mar{\'e}e}, \citenamefont {Grieneisen},\ and\ \citenamefont
  {Hogeweg}}]{maree2007cellular}%
  \BibitemOpen
  \bibfield  {author} {\bibinfo {author} {\bibfnamefont {A.~F.~M.}\
  \bibnamefont {Mar{\'e}e}}, \bibinfo {author} {\bibfnamefont {V.~A.}\
  \bibnamefont {Grieneisen}}, \ and\ \bibinfo {author} {\bibfnamefont
  {P.}~\bibnamefont {Hogeweg}},\ }\enquote {\bibinfo {title} {Single-cell-based
  models in biology and medicine},}\ \ (\bibinfo  {publisher} {Springer},\
  \bibinfo {year} {2007})\ Chap.\ \bibinfo {chapter} {The Cellular Potts Model
  and biophysical properties of cells, tissues and morphogenesis}, pp.\
  \bibinfo {pages} {107--136}\BibitemShut {NoStop}%
\bibitem [{\citenamefont {Shirinifard}\ \emph {et~al.}(2009)\citenamefont
  {Shirinifard}, \citenamefont {Gens}, \citenamefont {Zaitlen}, \citenamefont
  {Pop{\l}awski}, \citenamefont {Swat},\ and\ \citenamefont
  {Glazier}}]{shirinifard20093d}%
  \BibitemOpen
  \bibfield  {author} {\bibinfo {author} {\bibfnamefont {A.}~\bibnamefont
  {Shirinifard}}, \bibinfo {author} {\bibfnamefont {J.~S.}\ \bibnamefont
  {Gens}}, \bibinfo {author} {\bibfnamefont {B.~L.}\ \bibnamefont {Zaitlen}},
  \bibinfo {author} {\bibfnamefont {N.~J.}\ \bibnamefont {Pop{\l}awski}},
  \bibinfo {author} {\bibfnamefont {M.}~\bibnamefont {Swat}}, \ and\ \bibinfo
  {author} {\bibfnamefont {J.~A.}\ \bibnamefont {Glazier}},\ }\href@noop {}
  {\bibfield  {journal} {\bibinfo  {journal} {PloS One}\ }\textbf {\bibinfo
  {volume} {4}},\ \bibinfo {pages} {e7190} (\bibinfo {year}
  {2009})}\BibitemShut {NoStop}%
\bibitem [{\citenamefont {Chauviere}\ \emph {et~al.}(2012)\citenamefont
  {Chauviere}, \citenamefont {Hatzikirou}, \citenamefont {Kevrekidis},
  \citenamefont {Lowengrub},\ and\ \citenamefont {Cristini}}]{CHKLC2012}%
  \BibitemOpen
  \bibfield  {author} {\bibinfo {author} {\bibfnamefont {A.}~\bibnamefont
  {Chauviere}}, \bibinfo {author} {\bibfnamefont {H.}~\bibnamefont
  {Hatzikirou}}, \bibinfo {author} {\bibfnamefont {I.~G.}\ \bibnamefont
  {Kevrekidis}}, \bibinfo {author} {\bibfnamefont {J.~S.}\ \bibnamefont
  {Lowengrub}}, \ and\ \bibinfo {author} {\bibfnamefont {V.}~\bibnamefont
  {Cristini}},\ }\href@noop {} {\bibfield  {journal} {\bibinfo  {journal} {AIP
  Advances}\ }\textbf {\bibinfo {volume} {2}},\ \bibinfo {pages} {1} (\bibinfo
  {year} {2012})}\BibitemShut {NoStop}%
\bibitem [{\citenamefont {Marconi}\ and\ \citenamefont
  {Tarazona}(1999)}]{marconi1999dynamic}%
  \BibitemOpen
  \bibfield  {author} {\bibinfo {author} {\bibfnamefont {U.~M.~B.}\
  \bibnamefont {Marconi}}\ and\ \bibinfo {author} {\bibfnamefont
  {P.}~\bibnamefont {Tarazona}},\ }\href@noop {} {\bibfield  {journal}
  {\bibinfo  {journal} {J. Chem. Phys.}\ }\textbf {\bibinfo {volume} {110}},\
  \bibinfo {pages} {8032} (\bibinfo {year} {1999})}\BibitemShut {NoStop}%
\bibitem [{\citenamefont {Archer}\ and\ \citenamefont {Evans}(2004)}]{ArEv04}%
  \BibitemOpen
  \bibfield  {author} {\bibinfo {author} {\bibfnamefont {A.~J.}\ \bibnamefont
  {Archer}}\ and\ \bibinfo {author} {\bibfnamefont {R.}~\bibnamefont {Evans}},\
  }\href@noop {} {\bibfield  {journal} {\bibinfo  {journal} {J. Chem. Phys.}\
  }\textbf {\bibinfo {volume} {121}},\ \bibinfo {pages} {4246} (\bibinfo {year}
  {2004})}\BibitemShut {NoStop}%
\bibitem [{\citenamefont {Archer}\ and\ \citenamefont
  {Rauscher}(2004)}]{archer2004}%
  \BibitemOpen
  \bibfield  {author} {\bibinfo {author} {\bibfnamefont {A.~J.}\ \bibnamefont
  {Archer}}\ and\ \bibinfo {author} {\bibfnamefont {M.}~\bibnamefont
  {Rauscher}},\ }\href@noop {} {\bibfield  {journal} {\bibinfo  {journal} {J.
  Phys. A: Math. Gen.}\ }\textbf {\bibinfo {volume} {37}},\ \bibinfo {pages}
  {9325} (\bibinfo {year} {2004})}\BibitemShut {NoStop}%
\bibitem [{\citenamefont {Evans}(1979)}]{evans1979nature}%
  \BibitemOpen
  \bibfield  {author} {\bibinfo {author} {\bibfnamefont {R.}~\bibnamefont
  {Evans}},\ }\href@noop {} {\bibfield  {journal} {\bibinfo  {journal} {Adv.
  Phys.}\ }\textbf {\bibinfo {volume} {28}},\ \bibinfo {pages} {143} (\bibinfo
  {year} {1979})}\BibitemShut {NoStop}%
\bibitem [{\citenamefont {Evans}(1992)}]{Evans92}%
  \BibitemOpen
  \bibfield  {author} {\bibinfo {author} {\bibfnamefont {R.}~\bibnamefont
  {Evans}},\ }\href@noop {} {\emph {\bibinfo {title} {Fundamentals of
  Inhomogeneous Fluids}}}\ (\bibinfo  {publisher} {Dekker},\ \bibinfo {address}
  {New York},\ \bibinfo {year} {1992})\ Chap.~\bibinfo {chapter}
  {3}\BibitemShut {NoStop}%
\bibitem [{\citenamefont {Hansen}\ and\ \citenamefont
  {McDonald}(2013)}]{hansen2013theory}%
  \BibitemOpen
  \bibfield  {author} {\bibinfo {author} {\bibfnamefont {J.-P.}\ \bibnamefont
  {Hansen}}\ and\ \bibinfo {author} {\bibfnamefont {I.~R.}\ \bibnamefont
  {McDonald}},\ }\href@noop {} {\emph {\bibinfo {title} {Theory of Simple
  Liquids: With Applications to Soft Matter}}}\ (\bibinfo  {publisher}
  {Academic Press},\ \bibinfo {year} {2013})\BibitemShut {NoStop}%
\bibitem [{\citenamefont {Likos}(2001)}]{likos2001effective}%
  \BibitemOpen
  \bibfield  {author} {\bibinfo {author} {\bibfnamefont {C.~N.}\ \bibnamefont
  {Likos}},\ }\href@noop {} {\bibfield  {journal} {\bibinfo  {journal} {Phys.
  Rep.}\ }\textbf {\bibinfo {volume} {348}},\ \bibinfo {pages} {267} (\bibinfo
  {year} {2001})}\BibitemShut {NoStop}%
\bibitem [{\citenamefont {Dautenhahn}\ and\ \citenamefont
  {Hall}(1994)}]{DaHa94}%
  \BibitemOpen
  \bibfield  {author} {\bibinfo {author} {\bibfnamefont {J.}~\bibnamefont
  {Dautenhahn}}\ and\ \bibinfo {author} {\bibfnamefont {C.~K.}\ \bibnamefont
  {Hall}},\ }\href@noop {} {\bibfield  {journal} {\bibinfo  {journal}
  {Macromolecules}\ }\textbf {\bibinfo {volume} {27}},\ \bibinfo {pages} {5399}
  (\bibinfo {year} {1994})}\BibitemShut {NoStop}%
\bibitem [{\citenamefont {Likos}\ \emph {et~al.}(1998)\citenamefont {Likos},
  \citenamefont {L{\"o}wen}, \citenamefont {Watzlawek}, \citenamefont {Abbas},
  \citenamefont {Jucknischke}, \citenamefont {Allgaier},\ and\ \citenamefont
  {Richter}}]{likos:prl:98}%
  \BibitemOpen
  \bibfield  {author} {\bibinfo {author} {\bibfnamefont {C.~N.}\ \bibnamefont
  {Likos}}, \bibinfo {author} {\bibfnamefont {H.}~\bibnamefont {L{\"o}wen}},
  \bibinfo {author} {\bibfnamefont {M.}~\bibnamefont {Watzlawek}}, \bibinfo
  {author} {\bibfnamefont {B.}~\bibnamefont {Abbas}}, \bibinfo {author}
  {\bibfnamefont {O.}~\bibnamefont {Jucknischke}}, \bibinfo {author}
  {\bibfnamefont {J.}~\bibnamefont {Allgaier}}, \ and\ \bibinfo {author}
  {\bibfnamefont {D.}~\bibnamefont {Richter}},\ }\href@noop {} {\bibfield
  {journal} {\bibinfo  {journal} {Phys. Rev. Lett.}\ }\textbf {\bibinfo
  {volume} {80}},\ \bibinfo {pages} {4450} (\bibinfo {year}
  {1998})}\BibitemShut {NoStop}%
\bibitem [{\citenamefont {Louis}\ \emph {et~al.}(2000)\citenamefont {Louis},
  \citenamefont {Bolhuis}, \citenamefont {Hansen},\ and\ \citenamefont
  {Meijer}}]{LBHM00}%
  \BibitemOpen
  \bibfield  {author} {\bibinfo {author} {\bibfnamefont {A.~A.}\ \bibnamefont
  {Louis}}, \bibinfo {author} {\bibfnamefont {P.~G.}\ \bibnamefont {Bolhuis}},
  \bibinfo {author} {\bibfnamefont {J.-P.}\ \bibnamefont {Hansen}}, \ and\
  \bibinfo {author} {\bibfnamefont {E.~J.}\ \bibnamefont {Meijer}},\
  }\href@noop {} {\bibfield  {journal} {\bibinfo  {journal} {Phys. Rev. Lett.}\
  }\textbf {\bibinfo {volume} {85}},\ \bibinfo {pages} {2522} (\bibinfo {year}
  {2000})}\BibitemShut {NoStop}%
\bibitem [{\citenamefont {Dzubiella}\ \emph {et~al.}(2001)\citenamefont
  {Dzubiella}, \citenamefont {Jusufi}, \citenamefont {Likos}, \citenamefont
  {von Ferber}, \citenamefont {L{\"o}wen}, \citenamefont {Stellbrink},
  \citenamefont {Allgaier}, \citenamefont {Richter}, \citenamefont {Schofield},
  \citenamefont {Smith}, \citenamefont {Poon},\ and\ \citenamefont
  {Pusey}}]{Dzubiella_2001}%
  \BibitemOpen
  \bibfield  {author} {\bibinfo {author} {\bibfnamefont {J.}~\bibnamefont
  {Dzubiella}}, \bibinfo {author} {\bibfnamefont {A.}~\bibnamefont {Jusufi}},
  \bibinfo {author} {\bibfnamefont {C.~N.}\ \bibnamefont {Likos}}, \bibinfo
  {author} {\bibfnamefont {C.}~\bibnamefont {von Ferber}}, \bibinfo {author}
  {\bibfnamefont {H.}~\bibnamefont {L{\"o}wen}}, \bibinfo {author}
  {\bibfnamefont {J.}~\bibnamefont {Stellbrink}}, \bibinfo {author}
  {\bibfnamefont {J.}~\bibnamefont {Allgaier}}, \bibinfo {author}
  {\bibfnamefont {D.}~\bibnamefont {Richter}}, \bibinfo {author} {\bibfnamefont
  {A.~B.}\ \bibnamefont {Schofield}}, \bibinfo {author} {\bibfnamefont {P.~A.}\
  \bibnamefont {Smith}}, \bibinfo {author} {\bibfnamefont {W.~C.~K.}\
  \bibnamefont {Poon}}, \ and\ \bibinfo {author} {\bibfnamefont {P.~N.}\
  \bibnamefont {Pusey}},\ }\href@noop {} {\bibfield  {journal} {\bibinfo
  {journal} {Phys. Rev. E}\ }\textbf {\bibinfo {volume} {64}},\ \bibinfo
  {pages} {010401(R)} (\bibinfo {year} {2001})}\BibitemShut {NoStop}%
\bibitem [{\citenamefont {Gotze}\ \emph {et~al.}(2004)\citenamefont {Gotze},
  \citenamefont {Harreis},\ and\ \citenamefont {Likos}}]{GHL04}%
  \BibitemOpen
  \bibfield  {author} {\bibinfo {author} {\bibfnamefont {I.~O.}\ \bibnamefont
  {Gotze}}, \bibinfo {author} {\bibfnamefont {H.~M.}\ \bibnamefont {Harreis}},
  \ and\ \bibinfo {author} {\bibfnamefont {C.~N.}\ \bibnamefont {Likos}},\
  }\href@noop {} {\bibfield  {journal} {\bibinfo  {journal} {J. Chem. Phys.}\
  }\textbf {\bibinfo {volume} {120}},\ \bibinfo {pages} {7761} (\bibinfo {year}
  {2004})}\BibitemShut {NoStop}%
\bibitem [{\citenamefont {Mladek}\ \emph {et~al.}(2005)\citenamefont {Mladek},
  \citenamefont {Fernaud}, \citenamefont {Kahl},\ and\ \citenamefont
  {Neumann}}]{MFKN05}%
  \BibitemOpen
  \bibfield  {author} {\bibinfo {author} {\bibfnamefont {B.~M.}\ \bibnamefont
  {Mladek}}, \bibinfo {author} {\bibfnamefont {M.~J.}\ \bibnamefont {Fernaud}},
  \bibinfo {author} {\bibfnamefont {G.}~\bibnamefont {Kahl}}, \ and\ \bibinfo
  {author} {\bibfnamefont {M.}~\bibnamefont {Neumann}},\ }\href@noop {}
  {\bibfield  {journal} {\bibinfo  {journal} {Condens. Matter Phys.}\ }\textbf
  {\bibinfo {volume} {8}},\ \bibinfo {pages} {135} (\bibinfo {year}
  {2005})}\BibitemShut {NoStop}%
\bibitem [{\citenamefont {Likos}(2006)}]{Likos06}%
  \BibitemOpen
  \bibfield  {author} {\bibinfo {author} {\bibfnamefont {C.}~\bibnamefont
  {Likos}},\ }\href@noop {} {\bibfield  {journal} {\bibinfo  {journal} {Soft
  Matter}\ }\textbf {\bibinfo {volume} {2}},\ \bibinfo {pages} {478} (\bibinfo
  {year} {2006})}\BibitemShut {NoStop}%
\bibitem [{\citenamefont {Lenz}\ \emph {et~al.}(2012)\citenamefont {Lenz},
  \citenamefont {Blaak}, \citenamefont {Likos},\ and\ \citenamefont
  {Mladek}}]{LBLM12}%
  \BibitemOpen
  \bibfield  {author} {\bibinfo {author} {\bibfnamefont {D.~A.}\ \bibnamefont
  {Lenz}}, \bibinfo {author} {\bibfnamefont {R.}~\bibnamefont {Blaak}},
  \bibinfo {author} {\bibfnamefont {C.~N.}\ \bibnamefont {Likos}}, \ and\
  \bibinfo {author} {\bibfnamefont {B.~M.}\ \bibnamefont {Mladek}},\
  }\href@noop {} {\bibfield  {journal} {\bibinfo  {journal} {Phys. Rev. Lett.}\
  }\textbf {\bibinfo {volume} {109}},\ \bibinfo {pages} {228301} (\bibinfo
  {year} {2012})}\BibitemShut {NoStop}%
\bibitem [{\citenamefont {Archer}\ and\ \citenamefont
  {Evans}(2001)}]{archer2001binary}%
  \BibitemOpen
  \bibfield  {author} {\bibinfo {author} {\bibfnamefont {A.~J.}\ \bibnamefont
  {Archer}}\ and\ \bibinfo {author} {\bibfnamefont {R.}~\bibnamefont {Evans}},\
  }\href@noop {} {\bibfield  {journal} {\bibinfo  {journal} {Phys. Rev. E}\
  }\textbf {\bibinfo {volume} {64}},\ \bibinfo {pages} {041501} (\bibinfo
  {year} {2001})}\BibitemShut {NoStop}%
\bibitem [{\citenamefont {Archer}\ \emph {et~al.}(2004)\citenamefont {Archer},
  \citenamefont {Likos},\ and\ \citenamefont {Evans}}]{ALE04}%
  \BibitemOpen
  \bibfield  {author} {\bibinfo {author} {\bibfnamefont {A.~J.}\ \bibnamefont
  {Archer}}, \bibinfo {author} {\bibfnamefont {C.~N.}\ \bibnamefont {Likos}}, \
  and\ \bibinfo {author} {\bibfnamefont {R.}~\bibnamefont {Evans}},\
  }\href@noop {} {\bibfield  {journal} {\bibinfo  {journal} {J. Phys.: Cond.
  Mat.}\ }\textbf {\bibinfo {volume} {16}},\ \bibinfo {pages} {L297} (\bibinfo
  {year} {2004})}\BibitemShut {NoStop}%
\bibitem [{\citenamefont {Gotze}\ \emph {et~al.}(2006)\citenamefont {Gotze},
  \citenamefont {Archer},\ and\ \citenamefont {Likos}}]{GAL06}%
  \BibitemOpen
  \bibfield  {author} {\bibinfo {author} {\bibfnamefont {I.~O.}\ \bibnamefont
  {Gotze}}, \bibinfo {author} {\bibfnamefont {A.~J.}\ \bibnamefont {Archer}}, \
  and\ \bibinfo {author} {\bibfnamefont {C.~N.}\ \bibnamefont {Likos}},\
  }\href@noop {} {\bibfield  {journal} {\bibinfo  {journal} {J. Chem. Phys.}\
  }\textbf {\bibinfo {volume} {124}},\ \bibinfo {eid} {084901} (\bibinfo {year}
  {2006})}\BibitemShut {NoStop}%
\bibitem [{\citenamefont {Mladek}\ \emph {et~al.}(2006)\citenamefont {Mladek},
  \citenamefont {Gottwald}, \citenamefont {Kahl}, \citenamefont {Neumann},\
  and\ \citenamefont {Likos}}]{MGKNL06}%
  \BibitemOpen
  \bibfield  {author} {\bibinfo {author} {\bibfnamefont {B.~M.}\ \bibnamefont
  {Mladek}}, \bibinfo {author} {\bibfnamefont {D.}~\bibnamefont {Gottwald}},
  \bibinfo {author} {\bibfnamefont {G.}~\bibnamefont {Kahl}}, \bibinfo {author}
  {\bibfnamefont {M.}~\bibnamefont {Neumann}}, \ and\ \bibinfo {author}
  {\bibfnamefont {C.~N.}\ \bibnamefont {Likos}},\ }\href@noop {} {\bibfield
  {journal} {\bibinfo  {journal} {Phys. Rev. Lett.}\ }\textbf {\bibinfo
  {volume} {96}},\ \bibinfo {pages} {045701} (\bibinfo {year}
  {2006})}\BibitemShut {NoStop}%
\bibitem [{\citenamefont {Moreno}\ and\ \citenamefont {Likos}(2007)}]{MoLi07}%
  \BibitemOpen
  \bibfield  {author} {\bibinfo {author} {\bibfnamefont {A.~J.}\ \bibnamefont
  {Moreno}}\ and\ \bibinfo {author} {\bibfnamefont {C.~N.}\ \bibnamefont
  {Likos}},\ }\href@noop {} {\bibfield  {journal} {\bibinfo  {journal} {Phys.
  Rev. Lett.}\ }\textbf {\bibinfo {volume} {99}},\ \bibinfo {pages} {107801}
  (\bibinfo {year} {2007})}\BibitemShut {NoStop}%
\bibitem [{\citenamefont {Overduin}\ and\ \citenamefont
  {Likos}(2009{\natexlab{a}})}]{OvLi09b}%
  \BibitemOpen
  \bibfield  {author} {\bibinfo {author} {\bibfnamefont {S.~D.}\ \bibnamefont
  {Overduin}}\ and\ \bibinfo {author} {\bibfnamefont {C.~N.}\ \bibnamefont
  {Likos}},\ }\href@noop {} {\bibfield  {journal} {\bibinfo  {journal} {J.
  Chem. Phys.}\ }\textbf {\bibinfo {volume} {131}},\ \bibinfo {pages} {034902}
  (\bibinfo {year} {2009}{\natexlab{a}})}\BibitemShut {NoStop}%
\bibitem [{\citenamefont {Carta}\ \emph {et~al.}(2012)\citenamefont {Carta},
  \citenamefont {Pini}, \citenamefont {Parola},\ and\ \citenamefont
  {Reatto}}]{CPPR12}%
  \BibitemOpen
  \bibfield  {author} {\bibinfo {author} {\bibfnamefont {M.}~\bibnamefont
  {Carta}}, \bibinfo {author} {\bibfnamefont {D.}~\bibnamefont {Pini}},
  \bibinfo {author} {\bibfnamefont {A.}~\bibnamefont {Parola}}, \ and\ \bibinfo
  {author} {\bibfnamefont {L.}~\bibnamefont {Reatto}},\ }\href@noop {}
  {\bibfield  {journal} {\bibinfo  {journal} {J. Phys.: Condes. Matter}\
  }\textbf {\bibinfo {volume} {24}},\ \bibinfo {pages} {284106} (\bibinfo
  {year} {2012})}\BibitemShut {NoStop}%
\bibitem [{\citenamefont {Archer}\ \emph {et~al.}(2013)\citenamefont {Archer},
  \citenamefont {Rucklidge},\ and\ \citenamefont {Knobloch}}]{ARK13}%
  \BibitemOpen
  \bibfield  {author} {\bibinfo {author} {\bibfnamefont {A.~J.}\ \bibnamefont
  {Archer}}, \bibinfo {author} {\bibfnamefont {A.~M.}\ \bibnamefont
  {Rucklidge}}, \ and\ \bibinfo {author} {\bibfnamefont {E.}~\bibnamefont
  {Knobloch}},\ }\href@noop {} {\bibfield  {journal} {\bibinfo  {journal}
  {Phys. Rev. Lett.}\ }\textbf {\bibinfo {volume} {111}},\ \bibinfo {pages}
  {165501} (\bibinfo {year} {2013})}\BibitemShut {NoStop}%
\bibitem [{\citenamefont {Archer}\ \emph {et~al.}(2014)\citenamefont {Archer},
  \citenamefont {Walters}, \citenamefont {Thiele},\ and\ \citenamefont
  {Knobloch}}]{archer2014solidification}%
  \BibitemOpen
  \bibfield  {author} {\bibinfo {author} {\bibfnamefont {A.~J.}\ \bibnamefont
  {Archer}}, \bibinfo {author} {\bibfnamefont {M.~C.}\ \bibnamefont {Walters}},
  \bibinfo {author} {\bibfnamefont {U.}~\bibnamefont {Thiele}}, \ and\ \bibinfo
  {author} {\bibfnamefont {E.}~\bibnamefont {Knobloch}},\ }\href@noop {}
  {\bibfield  {journal} {\bibinfo  {journal} {Phys. Rev. E}\ }\textbf {\bibinfo
  {volume} {90}},\ \bibinfo {pages} {042404} (\bibinfo {year}
  {2014})}\BibitemShut {NoStop}%
\bibitem [{\citenamefont {O'Connor}\ \emph {et~al.}(2010)\citenamefont
  {O'Connor}, \citenamefont {Adams},\ and\ \citenamefont
  {Fairman}}]{o2010essentials}%
  \BibitemOpen
  \bibfield  {author} {\bibinfo {author} {\bibfnamefont {C.~M.}\ \bibnamefont
  {O'Connor}}, \bibinfo {author} {\bibfnamefont {J.~U.}\ \bibnamefont {Adams}},
  \ and\ \bibinfo {author} {\bibfnamefont {J.}~\bibnamefont {Fairman}},\
  }\href@noop {} {\bibfield  {journal} {\bibinfo  {journal} {Cambridge: NPG
  Education}\ } (\bibinfo {year} {2010})}\BibitemShut {NoStop}%
\bibitem [{\citenamefont {Mohr}\ \emph {et~al.}(2012)\citenamefont {Mohr},
  \citenamefont {Taylor},\ and\ \citenamefont {Newell}}]{mohr2012codata}%
  \BibitemOpen
  \bibfield  {author} {\bibinfo {author} {\bibfnamefont {P.~J.}\ \bibnamefont
  {Mohr}}, \bibinfo {author} {\bibfnamefont {B.~N.}\ \bibnamefont {Taylor}}, \
  and\ \bibinfo {author} {\bibfnamefont {D.~B.}\ \bibnamefont {Newell}},\
  }\href@noop {} {\bibfield  {journal} {\bibinfo  {journal} {Journal of
  Physical and Chemical Reference Data}\ }\textbf {\bibinfo {volume} {41}},\
  \bibinfo {pages} {043109} (\bibinfo {year} {2012})}\BibitemShut {NoStop}%
\bibitem [{\citenamefont {Hindmarsh}(1983)}]{hindmarsh1983odepack}%
  \BibitemOpen
  \bibfield  {author} {\bibinfo {author} {\bibfnamefont {A.~C.}\ \bibnamefont
  {Hindmarsh}},\ }\href@noop {} {\bibfield  {journal} {\bibinfo  {journal}
  {IMACS Transactions on Scientific Computation}\ }\textbf {\bibinfo {volume}
  {1}},\ \bibinfo {pages} {55} (\bibinfo {year} {1983})}\BibitemShut {NoStop}%
\bibitem [{\citenamefont {Hindmarsh}(2002)}]{hindmarsh2002serial}%
  \BibitemOpen
  \bibfield  {author} {\bibinfo {author} {\bibfnamefont {A.~C.}\ \bibnamefont
  {Hindmarsh}},\ }\href@noop {} {\bibfield  {journal} {\bibinfo  {journal}
  {URL: http://www.llnl.gov/CASC/odepack}\ } (\bibinfo {year}
  {2002})}\BibitemShut {NoStop}%
\bibitem [{\citenamefont {Archer}(2005)}]{archer2005dynamical}%
  \BibitemOpen
  \bibfield  {author} {\bibinfo {author} {\bibfnamefont {A.~J.}\ \bibnamefont
  {Archer}},\ }\href@noop {} {\bibfield  {journal} {\bibinfo  {journal} {J.
  Phys.: Condens. Matter}\ }\textbf {\bibinfo {volume} {17}},\ \bibinfo {pages}
  {1405} (\bibinfo {year} {2005})}\BibitemShut {NoStop}%
\bibitem [{\citenamefont {Derenzini}\ \emph {et~al.}(1998)\citenamefont
  {Derenzini}, \citenamefont {Trere}, \citenamefont {Pession}, \citenamefont
  {Montanaro}, \citenamefont {Sirri},\ and\ \citenamefont
  {Ochs}}]{derenzini1998nucleolar}%
  \BibitemOpen
  \bibfield  {author} {\bibinfo {author} {\bibfnamefont {M.}~\bibnamefont
  {Derenzini}}, \bibinfo {author} {\bibfnamefont {D.}~\bibnamefont {Trere}},
  \bibinfo {author} {\bibfnamefont {A.}~\bibnamefont {Pession}}, \bibinfo
  {author} {\bibfnamefont {L.}~\bibnamefont {Montanaro}}, \bibinfo {author}
  {\bibfnamefont {V.}~\bibnamefont {Sirri}}, \ and\ \bibinfo {author}
  {\bibfnamefont {R.~L.}\ \bibnamefont {Ochs}},\ }\href@noop {} {\bibfield
  {journal} {\bibinfo  {journal} {The American journal of pathology}\ }\textbf
  {\bibinfo {volume} {152}},\ \bibinfo {pages} {1291} (\bibinfo {year}
  {1998})}\BibitemShut {NoStop}%
\bibitem [{\citenamefont {Archer}\ \emph {et~al.}(2016)\citenamefont {Archer},
  \citenamefont {Walters}, \citenamefont {Thiele},\ and\ \citenamefont
  {Knobloch}}]{archer2016generation}%
  \BibitemOpen
  \bibfield  {author} {\bibinfo {author} {\bibfnamefont {A.~J.}\ \bibnamefont
  {Archer}}, \bibinfo {author} {\bibfnamefont {M.~C.}\ \bibnamefont {Walters}},
  \bibinfo {author} {\bibfnamefont {U.}~\bibnamefont {Thiele}}, \ and\ \bibinfo
  {author} {\bibfnamefont {E.}~\bibnamefont {Knobloch}},\ }in\ \href@noop {}
  {\emph {\bibinfo {booktitle} {Mathematical Challenges in a New Phase of
  Materials Science}}}\ (\bibinfo  {publisher} {Springer},\ \bibinfo {year}
  {2016})\ pp.\ \bibinfo {pages} {1--26}\BibitemShut {NoStop}%
\bibitem [{\citenamefont {Maru{\v{s}}i{\'c}}\ \emph {et~al.}(1994)\citenamefont
  {Maru{\v{s}}i{\'c}}, \citenamefont {Vuk-Pavlovic},\ and\ \citenamefont
  {Freyer}}]{maruvsic1994tumor}%
  \BibitemOpen
  \bibfield  {author} {\bibinfo {author} {\bibfnamefont {M.}~\bibnamefont
  {Maru{\v{s}}i{\'c}}}, \bibinfo {author} {\bibfnamefont {S.}~\bibnamefont
  {Vuk-Pavlovic}}, \ and\ \bibinfo {author} {\bibfnamefont {J.~P.}\
  \bibnamefont {Freyer}},\ }\href@noop {} {\bibfield  {journal} {\bibinfo
  {journal} {Bulletin of Mathematical Biology}\ }\textbf {\bibinfo {volume}
  {56}},\ \bibinfo {pages} {617} (\bibinfo {year} {1994})}\BibitemShut
  {NoStop}%
\bibitem [{\citenamefont {Archer}\ \emph {et~al.}(2002)\citenamefont {Archer},
  \citenamefont {Likos},\ and\ \citenamefont {Evans}}]{ALE02}%
  \BibitemOpen
  \bibfield  {author} {\bibinfo {author} {\bibfnamefont {A.~J.}\ \bibnamefont
  {Archer}}, \bibinfo {author} {\bibfnamefont {C.~N.}\ \bibnamefont {Likos}}, \
  and\ \bibinfo {author} {\bibfnamefont {R.}~\bibnamefont {Evans}},\
  }\href@noop {} {\bibfield  {journal} {\bibinfo  {journal} {J. Phys.: Cond.
  Mat.}\ }\textbf {\bibinfo {volume} {14}},\ \bibinfo {pages} {12031} (\bibinfo
  {year} {2002})}\BibitemShut {NoStop}%
\bibitem [{\citenamefont {Overduin}\ and\ \citenamefont
  {Likos}(2009{\natexlab{b}})}]{OvLi09}%
  \BibitemOpen
  \bibfield  {author} {\bibinfo {author} {\bibfnamefont {S.~D.}\ \bibnamefont
  {Overduin}}\ and\ \bibinfo {author} {\bibfnamefont {C.~N.}\ \bibnamefont
  {Likos}},\ }\href@noop {} {\bibfield  {journal} {\bibinfo  {journal}
  {Europhys. Lett.}\ }\textbf {\bibinfo {volume} {85}},\ \bibinfo {pages}
  {26003} (\bibinfo {year} {2009}{\natexlab{b}})}\BibitemShut {NoStop}%
\bibitem [{\citenamefont {Drasdo}\ \emph {et~al.}(2007)\citenamefont {Drasdo},
  \citenamefont {Hoehme},\ and\ \citenamefont {Block}}]{drasdo2007role}%
  \BibitemOpen
  \bibfield  {author} {\bibinfo {author} {\bibfnamefont {D.}~\bibnamefont
  {Drasdo}}, \bibinfo {author} {\bibfnamefont {S.}~\bibnamefont {Hoehme}}, \
  and\ \bibinfo {author} {\bibfnamefont {M.}~\bibnamefont {Block}},\
  }\href@noop {} {\bibfield  {journal} {\bibinfo  {journal} {J. Stat. Phys.}\
  }\textbf {\bibinfo {volume} {128}},\ \bibinfo {pages} {287} (\bibinfo {year}
  {2007})}\BibitemShut {NoStop}%
\bibitem [{\citenamefont {Werahera}\ \emph {et~al.}(2011)\citenamefont
  {Werahera}, \citenamefont {Glode}, \citenamefont {La~Rosa}, \citenamefont
  {Lucia}, \citenamefont {Crawford}, \citenamefont {Easterday}, \citenamefont
  {Sullivan}, \citenamefont {Sidhu}, \citenamefont {Genova},\ and\
  \citenamefont {Hedlund}}]{werahera2011proliferative}%
  \BibitemOpen
  \bibfield  {author} {\bibinfo {author} {\bibfnamefont {P.~N.}\ \bibnamefont
  {Werahera}}, \bibinfo {author} {\bibfnamefont {L.~M.}\ \bibnamefont {Glode}},
  \bibinfo {author} {\bibfnamefont {F.~G.}\ \bibnamefont {La~Rosa}}, \bibinfo
  {author} {\bibfnamefont {M.~S.}\ \bibnamefont {Lucia}}, \bibinfo {author}
  {\bibfnamefont {E.~D.}\ \bibnamefont {Crawford}}, \bibinfo {author}
  {\bibfnamefont {K.}~\bibnamefont {Easterday}}, \bibinfo {author}
  {\bibfnamefont {H.~T.}\ \bibnamefont {Sullivan}}, \bibinfo {author}
  {\bibfnamefont {R.~S.}\ \bibnamefont {Sidhu}}, \bibinfo {author}
  {\bibfnamefont {E.}~\bibnamefont {Genova}}, \ and\ \bibinfo {author}
  {\bibfnamefont {T.}~\bibnamefont {Hedlund}},\ }\href@noop {} {\bibfield
  {journal} {\bibinfo  {journal} {Prostate Cancer}\ }\textbf {\bibinfo {volume}
  {2011}} (\bibinfo {year} {2011})}\BibitemShut {NoStop}%
\bibitem [{\citenamefont {Sherwood}\ \emph {et~al.}(1992)\citenamefont
  {Sherwood}, \citenamefont {Stagnitti}, \citenamefont {Kokkinn},\ and\
  \citenamefont {Williams}}]{sherwood1992standard}%
  \BibitemOpen
  \bibfield  {author} {\bibinfo {author} {\bibfnamefont {J.~E.}\ \bibnamefont
  {Sherwood}}, \bibinfo {author} {\bibfnamefont {F.}~\bibnamefont {Stagnitti}},
  \bibinfo {author} {\bibfnamefont {M.~J.}\ \bibnamefont {Kokkinn}}, \ and\
  \bibinfo {author} {\bibfnamefont {W.~D.}\ \bibnamefont {Williams}},\
  }\href@noop {} {\bibfield  {journal} {\bibinfo  {journal} {International
  Journal of Salt Lake Research}\ }\textbf {\bibinfo {volume} {1}},\ \bibinfo
  {pages} {1} (\bibinfo {year} {1992})}\BibitemShut {NoStop}%
\bibitem [{\citenamefont {Ward}\ and\ \citenamefont
  {King}(1999{\natexlab{a}})}]{ward1999mathematical}%
  \BibitemOpen
  \bibfield  {author} {\bibinfo {author} {\bibfnamefont {J.~P.}\ \bibnamefont
  {Ward}}\ and\ \bibinfo {author} {\bibfnamefont {J.~R.}\ \bibnamefont
  {King}},\ }\href@noop {} {\bibfield  {journal} {\bibinfo  {journal}
  {Mathematical Medicine and Biology}\ }\textbf {\bibinfo {volume} {16}},\
  \bibinfo {pages} {171} (\bibinfo {year} {1999}{\natexlab{a}})}\BibitemShut
  {NoStop}%
\bibitem [{\citenamefont {Ward}\ and\ \citenamefont
  {King}(1999{\natexlab{b}})}]{ward1999math}%
  \BibitemOpen
  \bibfield  {author} {\bibinfo {author} {\bibfnamefont {J.~P.}\ \bibnamefont
  {Ward}}\ and\ \bibinfo {author} {\bibfnamefont {J.~R.}\ \bibnamefont
  {King}},\ }\href@noop {} {\bibfield  {journal} {\bibinfo  {journal}
  {Computational and Mathematical Methods in Medicine}\ }\textbf {\bibinfo
  {volume} {1}},\ \bibinfo {pages} {287} (\bibinfo {year}
  {1999}{\natexlab{b}})}\BibitemShut {NoStop}%
\end{thebibliography}

%

\end{document}